\documentclass[
12pt,tightenlines, amsmath, amssymb, nofootinbib, prd,
superscriptaddress, showpacs, preprintnumbers]{revtex4}
\newcommand{\beq}{\begin{equation}}
\newcommand{\eeq}{\end{equation}}
\newcommand{\ba}{\begin{array}{ccc}}
\newcommand{\ea}{\end{array}}
\newcommand{\nn}{\nonumber}
 \renewcommand{\d}{\partial}
\def\beqn{\begin{eqnarray}}
\def\eeqn{\end{eqnarray}}

\def\psit{\tilde{\psi}}
\def\Tr{ {\rm Tr} }
\def\<{\langle}
\def\>{\rangle}
\def\pp{{\langle\bar{\psi}\psi\rangle}_0}
\def\cond{ \frac{b g^2}{32 \pi^2 } G_{\mu \nu}^a G^{\mu \nu a}}
\def\GDG{\frac{g^2 G \tilde{G}}{32 \pi^2}}
\usepackage{graphicx}
\usepackage{amsmath}
\usepackage{amssymb}
\usepackage{psfrag}
\usepackage{epsfig}
\begin{document}

\title{\Large{{\bf $\theta-$ Parameter in 2 Color QCD\\at Finite Baryon and Isospin Density}}}
 \affiliation{Department of Physics and
Astronomy, University of British Columbia, Vancouver, BC, Canada,
V6T 1Z1}
 \author{ Max~A.~Metlitski}
 \email{mmetlits@phas.ubc.ca}
 \affiliation{Department of Physics and
Astronomy, University of British Columbia, Vancouver, BC, Canada,
V6T 1Z1}
 \author{ Ariel~R.~Zhitnitsky}
 \email{arz@phas.ubc.ca}
\affiliation{Department of Physics and
Astronomy, University of British Columbia, Vancouver, BC, Canada,
V6T 1Z1}

\date{\today }

\vfill
\begin{abstract}
We use 2-color QCD as a model to study the effects of simultaneous
presence of the so-called $\theta$ parameter, 
chemical potentials for baryon number, $\mu_B$ and for isospin
charge, $\mu_I$.  We pay special attention to
$\theta,~\mu_{B},~\mu_{I}$ dependence of different vacuum
condensates, including chiral and diquark condensates, as well as
the gluon condensate, $\langle\cond\rangle$, and the topological
susceptibility. We find that two phase transitions of the second
order will occur when $\theta$ relaxes from $\theta=2\pi$ to
$\theta=0$, if $\mu$ is of order of the pion mass, $ m_{\pi} $. We
demonstrate that the transition to the superfluid phase at $\theta
= \pi$ occurs at a much lower chemical potential than at $\theta =
0$. We also show that the strong $\theta$ dependence present near
$\theta = \pi$ in vacuum (Dashen's phenomenon), becomes smoothed
out in the superfluid phase.  Finally, we comment on the relevance
of this study for the real world with $N_c=3$.
\end{abstract}

\vfill

\maketitle

\section{Motivation}

 In this paper we investigate the behavior of
2-color QCD under the influence of three parameters: $\theta$,
$\mu_B$ and $\mu_I$. The main motivation for such a study is, of
course, the attempt to understand the cosmological phase
transition when $\theta$, being non-zero and large at the very
beginning of the phase transition, slowly relaxes to zero, as the
axion resolution of the strong CP problem suggests. Therefore, the
universe may undergo many QCD phase transitions when $\theta$
relaxes to zero. Another motivation is the attempt to understand
the complicated phase diagram of QCD as a function of  external
parameters $\theta$, $\mu_B$ and $\mu_I$. Finally, our study may
be of interests for the lattice community -- the determinant of
the Dirac operator for $N_c = 2$ is real when $\theta= \pi$ in the
presence of nonzero $\mu$. As we show, in this case the superfluid
phase is realized at a much lower chemical potential than at
$\theta=0$. This gives a unique chance to study the superfluid
phase on the lattice at a much smaller $\mu$ than would normally
be required.

To study all these problems in real 3 color QCD at finite $\mu_B$
is, of course, a very difficult task. To get some insight into
what might happen we
 shall use a controlled analytical method to study these questions in  the
non-physical (but nevertheless, very suggestive) $N_c=2$ theory.
We use
 the chiral effective Lagrangian approach to attack the problem.
 We shall determine the phase diagram in the
$\mu_B,\mu_I, \theta$ planes, various condensates and lowest lying
excitations. We expect that our approach is valid as long as all
external parameters $\mu_B, \mu_I$ and the quark mass, $m_q $, are
much smaller than $\Lambda_{QCD}$. We perform most of our
calculations for the case of two flavors $N_f = 2$ where the
algebra simplifies considerably.

One exciting effect that we find is that $\theta$ dependence of
the theory at fixed $\mu$ may become non-analytic. This is due to
the fact that the critical chemical potential for transition to
the superfluid phase varies with $\theta$. Therefore, a change of
$\theta$ might trigger a second-order phase transition,
accompanied by a discontinuity in the topological susceptibility
$\chi$. We also find that the strong $\theta$ dependence, present
near $\theta = \pi$ in vacuum, is washed out in the superfluid
phase. We expect that for equal quark masses a first order phase
transition (Dashen's phenomenon) will occur in the $N_c = 2$, $N_f
= 2$ theory at $\theta = \pi$, in the normal phase, but will
disappear in the superfluid phase.

We also find some interesting results, which appear even at
$\theta = 0$. Most importantly we compute the dependence of the
gluon condensate, $\langle\cond\rangle$, on the chemical
potential. The gluon condensate decreases with density near the
normal to superfluid phase transition, but, counter-intuitively,
increases for $m_\pi \ll \mu \ll \Lambda_{QCD}$.

We also  evaluate
 novel vacuum expectation values   which appear in the superfluid phase:
$\langle i u^T \gamma_0 C \gamma_5 \tau_2 d\rangle$ in the baryon
breaking phase and $\langle i \bar{u} \gamma_0 \gamma_5 d\rangle$
in the isospin breaking phase. These densities, being nonzero even at $\theta=0$, nonetheless
 have never been discussed  in the literature previously.
These densities, themselves, break
the baryon and isospin symmetries respectively, and so may be
considered as additional order parameters.

The presentation of our results is organized as follows.   In
section II,
  we introduce our notations for  the low energy effective Lagrangian. In section
  III,
  we introduce the $\theta$ parameter into the effective Lagrangian description.
  In section IV, we discuss the phase diagram of our theory in detail, computing the spectrum of lowest lying
  excitations, characterizing the phases in terms of chiral condensates and densities and paying special attention to
  physics near the point $\theta = \pi$.  In section V, we check that our results satisfy known Ward Identities supporting the self consistency of our approach.
   In section VI, we study the gluon condensate
  $\langle\cond\rangle$ as a function of $\mu$ and $\theta$. 
 In  Conclusion, we discuss the relevance of the obtained
  results for 3 color QCD and make some speculative remarks on evolution of the early universe
  during the QCD phase transition.
  In appendix I,
we  clarify some technical issues associated with  global
   aspects of the goldstone manifold.

\section{The Effective Theory at Finite $\mu_B$ and $\mu_I$ }

Two color QCD at zero chemical potential is invariant under
SU(2$N_f$) rotations in the chiral limit. This enhanced symmetry
(as compared to the SU($N_f)\times$SU($N_f)\times$U(1) of three
color QCD)  is manifest in the Lagrangian if we choose to
represent it in a basis of quarks $\psi$ and conjugate quarks
$\psit$ \cite{kogut1,kogut2}. For $N_f=2$ we use, \beq \Psi\equiv
\left(\begin{array}{c} u \\ d \\ \tilde u \\ \tilde d
  \end{array}\right)\equiv\left(\begin{array}{c} u_L \\ d_L \\
    \sigma_2\tau_2(u_R)^* \\ \sigma_2\tau_2(d_R)^*
  \end{array}\right)
\label{basis} \eeq where the Pauli matrices $\tau_2$ and
$\sigma_2$ act in colour and spin space respectively. We work in
Euclidean space and use the definitions, $\gamma_{\nu} =
\left(\begin{array}{cc} 0 & {\sigma_{\nu}}^{\dagger}\\\sigma_{\nu}
& 0\end{array}\right)$, $\gamma_5 = \left(\begin{array}{cc} -1 & 0
\\0 & 1 \end{array}\right)$, $\sigma_{\nu} = (-i, \sigma_k)$.
The microscopic Lagrangian then reads, \beq L = i \Psi^{\dagger}
\sigma_{\nu} D_{\nu} \Psi\eeq and possesses a symmetry, \beq
\label{symm} \Psi \rightarrow U \Psi, \,\,\, U \in SU(4)\eeq

The enhanced symmetry manifests itself in the low energy effective
theory through the manifold of goldstone modes associated with
spontaneous breaking of chiral symmetry, $SU(2 N_f) \rightarrow
Sp(2 N_f)$. In our case, $N_f = 2$, and the goldstone manifold is
SU(4)/Sp(4), corresponding to the condensation of $\Psi\Psi$
--- SU(4) flavor sextet. The fields on this manifold can be
represented by a 4$\times4$ antisymmetric unitary matrix $\Sigma$,
with $\det\Sigma=1$, that transforms  under (\ref{symm}) as, \beq
\Sigma \rightarrow U \Sigma U^T \eeq We parameterize the vacuum
manifold as, \beqn \label{Sigmac}\Sigma &=& U \Sigma_c U^T,
\,\,\,U \in SU(4), \,\,\,
\Sigma_c = \left(\begin{array}{cc} 0 & -1\\1 & 0\\
\end{array}\right)\eeqn

In what follows we use notations suggested in
\cite{Son:2000xc,Son:2000by,Splittorff:2000mm} for the description
of baryonic as well as isospin chemical potentials. In these
notations the baryon charge of the quark is 1/2, which comes from
$1/N_c$, so that the baryon (diquark in $N_c=2$) has baryon charge
1.
Thus, chemical potentials enter the microscopic Lagrangian as,
\beq  L = \bar{\psi} \gamma_{\nu} D_{\nu} \psi - \frac{1}{2}\mu_B
\bar{\psi}\gamma^0\psi - \frac{1}{2}\mu_I\bar{\psi}\gamma^0
\sigma^3\psi\eeq

In the basis of SU(4) spinors (\ref{basis}) the  baryon  and
isospin (third component) charge matrices in block-diagonal form
 are\cite{kogut1,kogut2,Son:2000xc,Son:2000by,Splittorff:2000mm},
 \beqn
\label{MBandI} B\equiv \frac12 \left(\begin{array}{cc} 1 & 0\\ 0 & -1\\
\end{array}\right) ,\ I\equiv \frac12 \left(\begin{array}{cc}
\sigma_3 &  0 \\  0 & -\sigma_3\\
\end{array}\right) \eeqn so that the Lagrangian reads,
\beq\label{Lmic}L = i \Psi^{\dagger} \sigma_{\nu} D_{\nu} \Psi-
\Psi^{\dagger}(\mu_B B + \mu_I I)\Psi\eeq

The effective Lagrangian for the field $\Sigma$ of goldstone modes
is determined by the symmetries inherited from the microscopic
two-color QCD Lagrangian. To lowest order in derivatives and at
zero quark mass the effective Lagrangian
is\cite{Splittorff:2000mm}, \beqn \label{L} {\cal L} &=&
\frac{F^2}{2} \Tr \nabla_{\nu} \Sigma \nabla_{\nu}
\Sigma^{\dagger}
\eeqn The $\mu$-dependence enters the effective Lagrangian through
the covariant extension of the derivative, \beqn\nn
 \partial_0 \Sigma & \to & \nabla_0 \Sigma = \partial_0 \Sigma -
\left[(\mu_BB+\mu_II)  \Sigma +  \Sigma (\mu_BB+\mu_II)^T\right], \,\, \nabla_i \Sigma = \d_i \Sigma \\
 \partial_0 \Sigma^\dagger & \to & \nabla_0\Sigma^{\dagger} = \partial_0 \Sigma^{\dagger} +
\left[(\mu_BB+\mu_II)  \Sigma +  \Sigma
(\mu_BB+\mu_II)^T\right]^\dagger, \,\, \nabla_i \Sigma = \d_i
\Sigma^{\dagger} \label{covar}\eeqn required by an extended local
gauge symmetry\cite{kogut1}. Therefore, to this order in chiral
perturbation theory, the Lagrangian at finite $\mu$ does not
require any extra phenomenological parameters beyond the pion
decay constant, $F$.
 This fact gives predictive power
to chiral perturbation theory at finite $\mu$. In using the
effective Lagrangian constructed above we must, of course, assume
that chiral symmetry for $N_c=N_f=2$ QCD is spontaneously broken.
 Since we have
regarded the hadronic modes as heavy, the theory is expected to be
valid only up to the mass of the lightest non-goldstone hadron.

 \section{The mass term and $\theta$ parameter}
 The mass term in the fundamental Lagrangian is defined as,
 \beq
 \label{1}
 L_m=m_u\bar{u}u+m_d\bar{d}d
 \eeq
 while the $\theta$ term in the fundamental Lagrangian is,
   \beq
 \label{1a}
 L_{\theta} =i\theta \cdot \frac{g^2G\tilde{G}}{32\pi^2}
 \eeq
 We keep $m_u\neq m_d$ on purpose: as is known $m_u=m_d$ is a very singular
 limit when one discusses $\theta$ dependence, see below. We
 would like to incorporate the $\theta$ dependence directly into
 the mass matrix. This can be achieved by performing a chiral
 rotation,
 \beq \label{gamma5}\psi \rightarrow e^{i \theta \gamma_5/2N_f} \psi\eeq
 With this field redefinition, the topological $\theta$ term in the Lagrangian
 disappears, due to the axial anomaly, and the mass term becomes,
 \beq \label{Lmicm}L_m = \bar{\psi}\frac{1+\gamma_5}{2}M^{\dagger}\psi +
 \bar{\psi}\frac{1-\gamma_5}{2}M\psi\eeq
 where the mass matrix M is,
\beqn M = e^{-i\theta/N_f}\left(\begin{array}{cc} m_u & 0\\0 & m_d
\end{array}\right)\eeqn
 In the basis of SU(4) spinors (\ref{basis}), the mass term
 becomes,
 \beq L_m = \frac{i}{2} \Psi^{T} {\cal M} \sigma_2 \tau_2 \Psi +
 h.c.\eeq
 where, in block-diagonal form,
 \beq
 {\cal M} = \left(\begin{array}{cc} 0 & M^{T}\\-M &
 0\end{array}\right)\eeq
The transformation properties of $L_m$ under (\ref{symm}) imply
that to lowest order, ${\cal M}$ enters the effective Lagrangian
as, \beq \label{mass} {\cal L}_{m}= -g Re\Tr\left({\cal M}
{\Sigma}\right), \eeq where the coefficient $g$ is determined by
the chiral condensate in the limit $m\rightarrow 0^+$, $\theta =
0$, $\mu_B = \mu_I = 0$ \cite{kogut2}, \beq g =
-\frac{{\langle\bar{\psi}\psi\rangle}_0}{2 N_f}\eeq as will be
confirmed below. In our notations the chiral condensate includes
the sum over all flavors, $\langle \bar{\psi}\psi\rangle =
\sum_f{\langle\bar{\psi}_f\psi_f\rangle}$.

The chiral effective Lagrangian incorporating the effects of
$\mu_B, \mu_I, \theta$ and non-zero quark masses, thus becomes,
\beq \label{Lch} {\cal L} = \frac{F^2}{2} \Tr \nabla_{\nu} \Sigma
\nabla_{\nu} \Sigma^{\dagger} - g Re\Tr\left({\cal M}
{\Sigma}\right)\eeq We shall use the Lagrangian (\ref{Lch}) for
the rest of this work.

So far our discussion easily generalizes to arbitrary $N_f$.
However, significant algebraic simplification can be obtained by
considering $N_f = 2$. Indeed, for $N_f = 2$, the effective
Lagrangian (\ref{Lch}) with $m_u \neq m_d$, $\theta \neq 0$, can
be reduced to the same Lagrangian but with $m_u' = m_d'$ and
$\theta' = 0$. This is achieved, by performing an $SU(4)$ (more
specifically $SU(2)_A$) rotation, \beq \label{rot} \Sigma = U_0
\tilde{\Sigma} {U_0}^T\eeq with the particular choice of, \beqn
\nn U_0 &=& \left(\begin{array}{cc} L & 0\\0
& R^*\end{array}\right), \,\,\,L = R^* = e^{i \alpha \sigma^3/2}\\
\cos{\alpha} &=& \frac{(m_u + m_d)
\cos(\theta/2)}{\sqrt{(m_u+m_d)^2 \cos^2(\theta/2) + (m_u-m_d)^2
\sin^2(\theta/2)}}\nn\\
\sin\alpha &=&  \frac{(m_u - m_d)
\sin(\theta/2)}{\sqrt{(m_u+m_d)^2 \cos^2(\theta/2) + (m_u-m_d)^2
\sin^2(\theta/2)}}\label{U0}\eeqn Our parameter $\alpha$ is
related to the commonly used Witten's variables $\phi_u,
\phi_d$\cite{Witten:1980sp}, via, \beqn \phi_u = \theta/2 -
\alpha, \quad \phi_d = \theta/2 + \alpha\\ \phi_u + \phi_d =
\theta, \quad m_u \sin \phi_u = m_d \sin \phi_d\eeqn After such a
transformation, the Lagrangian (\ref{Lch}) takes the form, \beq
\label{Ls}{\cal L} = \frac{F^2}{2} \Tr \nabla_{\nu} \tilde{\Sigma}
\nabla_{\nu} \tilde{\Sigma}^{\dagger} - g m(\theta)
Re\Tr\left({\cal M}_0\tilde{\Sigma}\right)\eeq with, \beqn
\label{m}{\cal M}_0 &=& \left(\begin{array}{cc} 0 & 1\\-1 &
0\end{array}\right),\,\,\, m(\theta) =
\frac{1}{2}\left({(m_u+m_d)^2\cos^2(\theta/2) +
(m_u-m_d)^2\sin^2(\theta/2)}\right)^{\frac12}\eeqn The detailed
explanation of this reduction, which along the way clarifies
certain global properties of the vacuum manifold, is presented in
apppendix I. Technically, the simplification is due to
pseudo-reality of $SU(N_f = 2)$ (see also section IVC for a more quantitative
discussion).


\section{Phase Diagram}
\subsection{Vacuum Alignment and Spectrum}
Our next step is to find the classical minimum of the effective
Lagrangian (\ref{Lch}) to determine the phase diagram, pattern of
spontaneous symmetry breaking and, subsequently, the spectrum of
excitations. For arbitrary $N_f$, quark masses, $\theta$ and
chemical potentials this is a non-trivial algebraic problem.
However, as was shown in the previous section, for $N_f = 2$, the
effective Lagrangian reduces to the form (\ref{Ls}), which was
already analyzed in \cite{Splittorff:2000mm}. Thus, we may
immediately read off all quantities of interest.

First, let's study the phase diagram for fixed $m_u, m_d$,
$\theta$. To get acquainted with our theory, let's begin with the
trivial environment $\mu_B = \mu_I = 0$. The effective Lagrangian
possesses an $Sp(4)$ symmetry at this point. The classical minimum
is given by, \beq \langle\tilde{\Sigma}\rangle = \Sigma_c\eeq The
$Sp(4)$ symmetry is unbroken. The low-lying excitations are a
quintet of pseudo-goldstones (3 pions and 2 diquarks), with
dispersions,
\beqn\label{dispv} E &=& \sqrt{{\bf p}^2 + m_{\pi}^2(\theta)}\\
m^2_{\pi}(\theta) &=& \frac{g m(\theta)}{F^2} = \frac{m(\theta)
|\pp|}{4 F^2}\eeqn The pseudo-Goldstone mass $m_{\pi}$ acquires a
dependence on $\theta$ through the effective quark mass parameter
$m(\theta)$ (\ref{m}) (this $\theta$ dependence is implicitly
implied in all formulas below, unless otherwise stated). As we
shall see, the whole phase diagram turns out to be determined by
the parameter $m_{\pi}(\theta)$. We note that $m_{\pi}(\theta)$
reaches its maximum at $\theta = 0$ and minimum at $\theta = \pi$.
Moreover, for $m_u = m_d$, $\theta = \pi$,  $m_\pi$ vanishes to
first order in $M$.

We note that strong $P$ and $CP$ symmetries are explicitly broken
in the system with $\theta \neq 0$. So at $\theta \neq 0$, the
pions (diquarks) are no longer pure pseudoscalars (scalars). This
will become particularly clear when we discuss Bose-condensates of
our goldstones in the superfluid phase.

Now let's turn on chemical potentials. For $\mu_B \neq 0$, $\mu_I
\neq 0$, the symmetry of the problem is broken to $U(1)_B \times
U(1)_I$.\footnote{If only one of the chemical potentials is turned
on, say $\mu_I = 0$, $\mu_B \neq 0$, then the symmetry is
actually, $SU(2)_V \times U(1)_B$.} We introduce the following
notations to describe vacuum alignment of $\Sigma$ at finite
chemical potentials, \beq \Sigma_B = \left(\begin{array}{cc}
\sigma_2 & 0\\0 & \sigma_2\\\end{array}\right),\quad\quad \Sigma_I
= i \left(\begin{array}{cc} 0 & \sigma_1\\-\sigma_1 &
0\\\end{array}\right)\eeq As is known\cite{Splittorff:2000mm},
there are 3 distinct phases in the ($\mu_B, \mu_I$)
plane,\footnote{In the original paper\cite{Splittorff:2000mm}, a
certain physically reasonable ansatz was taken for the classical
static minimum $\langle \tilde{\Sigma}\rangle$ of (\ref{Lch}). It
was shown that this ansatz is, indeed, a local minimum, and the
authors assumed that this minimum is also global. We note, that
using the explicit parametrization of the vacuum manifold
presented in Appendix I, it is possible to prove that the ansatz
is, indeed, a global minimum.}

\vspace{0.5cm}

I. \quad Normal Phase (N): \quad$|\mu_B| < m_{\pi}(\theta),\,
|\mu_I| < m_{\pi}(\theta)$ \vspace{0.3cm}\beq
\langle\tilde{\Sigma}\rangle = \Sigma_c\eeq \hspace{1.4cm}Symmetry
breaking: $\hspace{1cm} U(1)_B \times U(1)_I \rightarrow U(1)_B
\times U(1)_I$

\vspace{0.3cm}

\hspace{0.7cm} Spectrum: \beqn
q^{\pm}\hspace{2cm} E& = &  \sqrt{{\bf p}^2+m^2_{\pi}}\pm \mu_B  \nn\\
\pi^0 \hspace{2cm} E & = & \sqrt{{\bf p}^2+m^2_{\pi}} \nn \\
\pi^{\pm}  \hspace{2cm} E & = & \sqrt{{\bf p}^2+m^2_{\pi}} \pm \mu_I \nn \\
\label{SN} \eeqn

\vspace{0.3cm}

II. \quad Baryon Phase (B): \quad$|\mu_B| > m_{\pi}(\theta),\,
|\mu_I| < |\mu_B|$ \vspace{0.25cm}\beq
\langle\tilde{\Sigma}\rangle = \frac{m^2_{\pi}}{\mu_B^2}
\,\Sigma_c + {\left(1 -
\frac{m^4_{\pi}}{\mu_B^4}\right)}^{\frac12}\,
\Sigma_B\eeq\hspace{1.4cm}Symmetry breaking: $\hspace{1cm} U(1)_B
\times U(1)_I \rightarrow U(1)_I$

\vspace{0.3cm}

\hspace{0.7cm} Spectrum: \beqn \tilde{q}^{\pm}\hspace{1.8cm} E^2 &
= & {\bf p}^2 + \frac{1}{2} \mu^2_B \left(1+ 3
\frac{m^4_{\pi}}{\mu^4_B}\right) \pm \mu_B \left(4 {\bf p}^2
\frac{m^4_{\pi}}{\mu^4_B} + \frac{1}{4} \mu^2_B \left(
1+3\frac{m^4_{\pi}}{\mu^4_B}\right)^2\right)^{\frac12}\nn\\
\pi^0 \hspace{2cm} E & = & \sqrt{{\bf p}^2+\mu^2_B} \nn \\
\pi^{\pm}  \hspace{2cm} E & = & \sqrt{{\bf p}^2+\mu^2_B} \pm \mu_I \nn \\
\label{SB} \eeqn \vspace{0.1cm}

III. \quad Isospin Phase (I): \quad$|\mu_I| > m_{\pi}(\theta),\,
|\mu_B| < |\mu_I|$ \vspace{0.25cm}\beq
\langle\tilde{\Sigma}\rangle = \frac{m^2_{\pi}}{\mu^2_I}
\,\Sigma_c + {\left(1 -
\frac{m^4_{\pi}}{\mu_I^4}\right)}^{\frac12}\,
\Sigma_I\eeq\hspace{1.4cm}Symmetry breaking: $\hspace{1cm} U(1)_B
\times U(1)_I \rightarrow U(1)_B$

\vspace{0.3cm}

\hspace{0.7cm} Spectrum: Same as for B Phase, but with $q^{\pm}
\leftrightarrow \pi^{\pm}$ and $\mu_B \leftrightarrow \mu_I$.

\vspace{0.3cm}

The phase transition between $N$ phase and $B$ phase, as well as
$N$ phase and $I$ phase is second order, whereas the phase
transition between $B$ phase and $I$ phase is first order. As
noted in \cite{Splittorff:2000mm}, the symmetry of the phase
diagram/spectrum, with respect to $\mu_B \leftrightarrow \mu_I$ is
a direct consequence of the symmetry of the microscopic theory,
$d_L \leftrightarrow -\tilde{d}_R$, $d_R \leftrightarrow
\tilde{d}_L$, $\mu_B \leftrightarrow \mu_I$.

Thus, the phase diagram in the ($\mu_B$, $\mu_I$) plane looks the
same at $\theta \neq 0$ as at $\theta = 0$, with the important
replacement, $m^2_\pi \rightarrow m^2_\pi(\theta)$. This is a very
natural conclusion. Indeed, at $\theta \neq 0$ diquarks (pions)
still carry baryon (isospin) number. Hence, their energy is
lowered at finite baryon (isospin) chemical potential. As soon as
$\mu_B$ ($\mu_I$) reaches the vacuum diquark (pion) mass
$m_\pi(\theta)$, Bose-condensation occurs leading to spontaneous
breaking of $U(1)_B$ ($U(1)_I$) symmetry.

Quantitatively, the $\theta$ dependence of the Goldstone mass
$m_\pi(\theta)$ implies that the transition to superfluid phase is
shifted to a smaller chemical potential $\mu_B$, $\mu_I$, compared
to $\theta = 0$. In the limiting case, when $m_u = m_d$ and
$\theta = \pi$, the transition occurs in the vicinity of $\mu = 0$
(see Section IVC for a more precise discussion). For physical
values, $m_d = 7 MeV$, $m_u = 4 MeV$, the transition at $\theta =
\pi$ occurs at $\mu = \left(\frac{m_d - m_u}{m_d +
m_u}\right)^{\frac12}m_\pi(0) \sim 70 MeV$.

\subsection{Chiral Condensates and Densities}

In section IVA, we have established the phase diagram of $N_c =
2$, $N_f = 2$ QCD at finite $\mu_B$, $\mu_I$ and $\theta$. In this
section, we wish to characterize this phase diagram in terms of
chiral condensates and densities. Similar computations have been
performed\cite{kogut2,Splittorff:2000mm} at $\theta = 0$. However,
we evaluate a wider range of expectation values and find some
condensates that are non-zero even at $\theta = 0$, which have not
been discussed in the original papers. As expected, we also find
new condensates at $\theta \neq 0$.

We follow the standard procedure for computing microscopic
condensates from the effective Lagrangian. We start from a
slightly generalized version of the microscopic Lagrangian
(\ref{Lmic}) together with the mass term (\ref{Lmicm}), \beq L = i
\Psi^{\dagger} \sigma_{\nu} D_{\nu} \Psi- \Psi^{\dagger}T\Psi +
\frac{i}{2} \left(\Psi^{T} J \sigma_2 \tau_2 \Psi - \Psi^{\dagger}
{J}^{\dagger} \sigma_2 \tau_2 \Psi^*\right)\eeq

Here the hermitian, traceless matrix $T$ incorporates the chemical
potentials for all $15$ charges associated with the $SU(4)$
symmetry, and the chiral condensate source ${J}$ is an arbitrary,
antisymmetric matrix (12 real components). We may express $T$ and
$J$ in terms of a basis, \beqn T = t_A \lambda_A, \quad J = j_a
X_a, \quad \lambda_A = \lambda_A^{\dagger}, \,\, \Tr(\lambda_A) =
0,\,\, {X_a}^T = - X_a, \,\, t_A, j_a \in {\mathbb R}\eeqn

Differentiating the vacuum free energy density $\cal F$, we obtain
our condensates and charge densities: \beqn\nn\frac{\d {\cal
F}}{\d t_A} &=& -
\langle\Psi^{\dagger}\lambda_A\Psi\rangle\\\frac{\d {\cal F}}{\d
j_a} &=& \frac{i}{2}\langle{\Psi^{T}X_a \sigma_2 \tau_2 \Psi -
\Psi^{\dagger} {{X}^{\dagger}}_a \sigma_2 \tau_2 \Psi^*}\rangle
\label{dF}\eeqn The relations (\ref{dF}) hold for any $T$, $J$,
however, we will apply them when the derivatives and expectation
values are evaluated at physical parameters, $T = T_0 = \mu_B B +
\mu_I I$ and $J = {\cal M}$.

In the effective theory, the sources $T$ and $J$ are incorporated
by replacing $T_0 \rightarrow T$ in the covariant derivative
(\ref{covar}), and ${\cal M} \rightarrow J$ in the mass term
(\ref{mass}). The condensates (\ref{dF}), thus, become,
\beqn\nn\frac{\d {\cal F}}{\d t_A} &=& F^2
\langle\Tr\left(\Sigma^{\dagger} \lambda_A \nabla_0 \Sigma -
\nabla_0 \Sigma^{\dagger} \lambda_A \Sigma\right)\rangle\\\frac{\d
{\cal F}}{\d j_a} &=& -g \langle Re\Tr\left(X_a
\Sigma\right)\rangle\label{dFeff}\eeqn

It remains to evaluate the expressions (\ref{dFeff}), with
$\Sigma$ given by the time-independent, classical minimum of the
effective Lagrangian (\ref{Lch}). We must remember that to
simplify algebra we expressed, $\Sigma = U_0 \tilde{\Sigma}
{U_0}^T$, with $U_0$ given by (\ref{U0}). We should also remember
that we incorporated $\theta$ dependence into the mass-matrix by a
chiral rotation (\ref{gamma5}) of the quark fields. The
condensates and densities, expressed in terms of the original
quark fields, are listed in Tables 1,2. We define the charge
conjugation matrix $C = \gamma_0 \gamma_2 \gamma_5$. We also
introduce the parameter $\lambda(\theta)$ in Tables 1,2,\beqn
\lambda(\theta) = \left\{ \begin{array}{cl}
1 &\quad \textrm{Normal Phase}\\
\frac{m^2_\pi(\theta)}{\mu^2_B} &\quad \textrm{Baryon Phase}\\
\frac{m^2_\pi(\theta)}{\mu^2_I} &\quad \textrm{Isospin Phase}
\end{array} \right.
\eeqn which obtains its $\theta$ dependence through
$m^2_\pi(\theta)$.

\begin{table}[!p]
\caption{Chiral condensates in $N_c = N_f = 2$ QCD at finite
$\theta$} \vspace{0.25cm}
\begin{tabular}{|c|c|c|c|}
\hline \rule{0pt}{4ex}Condensate$/\pp$ & N Phase ($\lambda = 1$) &
B Phase ($\lambda = \frac{m^2_\pi(\theta)}{\mu^2_B}$)&
I Phase ($\lambda = \frac{m^2_\pi(\theta)}{\mu^2_I}$)\\
\hline \rule{0pt}{3.5ex}$i u^{T}C \gamma_5\tau_2 d$ & $0$ &
$-\frac12 \cos(\frac{\theta}{2}) \left(1
-\lambda^2\right)^{\frac12}$ & $0$\\
\hline \rule{0pt}{3.5ex}$u^{T}C \tau_2 d$ & $0$ & $-\frac12
\sin(\frac{\theta}{2}) \left(1 -
\lambda^2\right)^{\frac12}$ & $0$\\
\hline \rule{0pt}{3.5ex}$i \bar{u} \gamma_5 d$ & $0$ & $0$
&$-\frac12 \cos(\frac{\theta}{2}) \left(1 -
\lambda^2\right)^{\frac12}$\\
\hline \rule{0pt}{3.5ex}$\bar{u} d$ & $0$ & $0$ &$-\frac12
\sin(\frac{\theta}{2}) \left(1
- \lambda^2\right)^{\frac12}$\\
\hline \rule{0pt}{3.5ex}$\bar{u} u$ &
\multicolumn{3}{|c|}{$\,\,\,\,\,\frac{1}{2}
\lambda \cos(\frac{\theta}{2} -\alpha)$}\\
\hline \rule{0pt}{3.5ex}$i \bar{u} \gamma_5 u$ &
\multicolumn{3}{|c|}{ $-\frac{1}{2} \lambda\sin(\frac{\theta}{2} -
\alpha)$}\\\hline\rule{0pt}{3.5ex}$\bar{d} d$ &
\multicolumn{3}{|c|}{$\,\,\,\,\,\frac{1}{2}
\lambda\cos(\frac{\theta}{2} + \alpha)$}
\\\hline\rule{0pt}{3.5ex}
$i\bar{d} \gamma_5 d$ & \multicolumn{3}{|c|}{$-\frac{1}{2}
\lambda\sin(\frac{\theta}{2} + \alpha)$} \\\hline
\end{tabular}
\end{table}

\begin{table}[!p]
\caption{Densities in $N_c = N_f = 2$ QCD at finite $\theta$}
\vspace{0.25cm}
\begin{tabular}{|c|c|c|c|}
\hline \rule{0pt}{4ex}Density & N Phase ($\lambda = 1$) & B Phase
($\lambda = \frac{m^2_\pi(\theta)}{\mu^2_B}$)&
I Phase ($\lambda = \frac{m^2_\pi(\theta)}{\mu^2_I}$)\\
\hline \rule{0pt}{3.5ex} $\frac{1}{2} \bar{\psi}\gamma_0 \psi$ &
$0$ & $4
F^2 \mu_B (1-\lambda^2)$ & 0\\
\hline \rule{0pt}{3.5ex} $i u^T \gamma_0 C \gamma_5 \tau_2 d$ &
$0$ & $-4 F^2 \mu_B \lambda \left(1-\lambda^2\right)^{\frac12}
\cos(\alpha)$ & $0$\\
\hline \rule{0pt}{3.5ex} $u^T \gamma_0 C \tau_2 d$ & $0$ & $-4 F^2
\mu_B
\lambda \left(1-\lambda^2\right)^{\frac12}\sin(\alpha)$ & $0$\\
\hline \rule{0pt}{3.5ex} $\frac{1}{2}\bar{\psi}\gamma_0 \sigma_3
\psi$ & $0$ & $0$ & $4 F^2 \mu_I \left(1-\lambda^2\right)$\\
\hline \rule{0pt}{3.5ex} $i \bar{u} \gamma_0 \gamma_5 d$ & $0$ &
$0$ &
$4 F^2 \mu_I \lambda \left(1 -\lambda^2\right)^{\frac12} \cos(\alpha)$\\
\hline \rule{0pt}{3.5ex} $\bar{u} \gamma_0  d$ & $0$ & $0$ &
$4 F^2 \mu_I \lambda \left(1 -\lambda^2\right)^{\frac12} \sin(\alpha)$\\
\hline \rule{0pt}{3.5ex} $u^T \gamma_0 C \tau_2 u$ & $0$ & $0$ &
$0$\\
\hline \rule{0pt}{3.5ex} $d^T \gamma_0 C \tau_2 d$ & $0$ & $0$ &
$0$\\
\hline
\end{tabular}
\vspace{5cm}
\end{table}

We can now see, how our phase diagram is described in terms of
condensates and charge densities. First, let's gauge our intuition
by considering the Normal phase. At $\theta = 0$, the parameter
$\alpha$ of eq.\,(\ref{U0}) is $0$, and the only condensates
(Table 1) are $\langle\bar{u} u\rangle = \langle \bar{d}
d\rangle$. At non-zero $\theta$, we also get condensates $\langle
i\bar{u} \gamma^5 u\rangle$, $\langle i \bar{d} \gamma_5
d\rangle$, while $\langle \bar{u} u\rangle$, $\langle \bar{d} d
\rangle$ get depleted. The appearance of $P$ and $CP$ odd
condensates $\langle i \bar{u} \gamma_5 u\rangle$, $\langle i
\bar{d}  \gamma_5 d\rangle$ is a direct consequence of explicit
$P$ and $CP$ breaking by the $\theta$ term. Finally, for $\theta
\neq 0$, $m_u \neq m_d$, the parameter $\alpha \neq 0$, and we see
explicit effects of isospin symmetry breaking: $\langle \bar{u}
u\rangle \neq \langle \bar{d} d \rangle$ (correspondingly for $P$
odd condensates). Such effects are absent to lowest order in $M$
at $\theta = 0$. As expected, all charge densities (Table 2) in
the Normal phase vanish.

Let's now see what happens in superfluid phases. At $\theta = 0$,
the Baryon phase is characterized by a scalar diquark condensate
$\langle i u^{T}C \gamma_5\tau_2 d\rangle$, which breaks the
$U(1)_B$ symmetry. The Isospin phase is characterized by a
pseudo-scalar pion condensate $\langle i \bar{u} \gamma_5 d
\rangle$, which breaks the $U(1)_I$ symmetry. These condensates
appear at the expense of depleting $\langle \bar{\psi} \psi
\rangle$. As expected, at finite $\theta$, $U(1)_B$ and $U(1)_I$
violating condensates of opposite parity also appear: $\langle
u^{T}C \tau_2 d\rangle$ in $B$ phase and $\langle \bar{u}
d\rangle$ in $I$ phase. 

The Baryon and Isospin phases also carry non-vanishing $SU(4)$
charge densities. The I phase, is characterized by the isospin
density, \beq \label{nI}n_I = \frac12 \langle \bar{\psi} \gamma_0
\sigma_3 \psi \rangle = 4 F^2 \mu_I \left(1 - \frac{m^4_\pi
(\theta)}{\mu^4_I}\right)\eeq This is precisely the density, which
one expects to induce by applying an isospin chemical potential
$\mu_I$. At $\theta =0$ it   coincides with the previous results
\cite{Son:2000by},\cite{Splittorff:2000mm}. In addition, we also
obtain the following axial charge density,

\beq \label{newdensI}n_A = \langle i \bar{u} \gamma_0 \gamma_5
d\rangle = 4 F^2 \mu_I \frac{m^2_\pi(\theta)}{\mu^2_I} \left(1
-\frac{m^4_\pi(\theta)}{\mu^4_I}\right)^{\frac12}
\cos\alpha(\theta)\eeq which has not been discussed previously in
the literature even at $\theta=0$. This is the axial charge
density, corresponding to off-diagonal generators of the $SU(2)_A$
group, which is both spontaneously and explicitly broken. Note
that the axial charge density (\ref{newdensI}) does not vanish
already at $\theta = 0$. Thus, we for now concentrate on $\theta =
0$, and hence $\alpha(\theta=0) = 0$, to better understand the
physical nature of this new density (\ref{newdensI}). For
simplicity, we take $|\mu_B| < m_\pi $.

The   density $n_A$ spontaneously breaks the $U(1)_I$ symmetry
and, hence, may be considered as an order parameter alongside the
pion condensate, \beq \label{pion}\langle \pi^- \rangle = \langle
i \bar{u} \gamma_5 d\rangle = -\frac12 \pp \left(1 -
\frac{m^4_\pi}{\mu^4_I}\right)^{\frac12}\eeq There was no explicit
chemical potential conjugate to $n_A$ in the Lagrangian - once
$U(1)_I$ is already spontaneously broken by $\langle \pi \rangle$,
$n_A$ is induced automatically. The quantitative behaviour of
these two order parameters is somewhat different. The pion
condensate monotonically increases with $\mu_I$ after the Normal
to Isospin phase transition, and $\langle \pi \rangle \to
-\frac{1}{2} \pp$ for $\mu_I \gg m_\pi$. On the other hand, the
new charge density $n_A$ first increases after the phase
transition, reaches a peak at $\mu_I = 3^{1/4} m_\pi$, and then
decreases to $0$ for $\mu \gg m_\pi$. Of course, we always
consider only $\mu_I, \mu_B \ll \Lambda_{QCD}$.

One can understand the appearance of a new condensate $n_A =
\langle i \bar{u} \gamma_0 \gamma_5 d\rangle$ in the following
simple way. We are in the phase where the isospin density, $n_I
\sim \langle \bar{u} \gamma_0 u \rangle -\langle \bar{d} \gamma_0
d \rangle$, as well as the condensate, $\<\pi^-\>\sim
 \langle i \bar{u} \gamma_5 d\rangle$, do not vanish. This implies that our ground state can be
 understood as a coherent superposition of an infinitely large number of $\pi^-$ mesons.
 We expect that we do not disturb the ground state of the system by adding one of these $\pi^-$ mesons.
 On the other hand, we can relate the matrix element with an extra $\pi^-$ meson to the matrix element without the
 $\pi^-$
 using the standard PCAC technique, $\< A| O|B \pi\>\sim
 i\< A| [O, Q^5]|B \>$. The coefficient of proportionality
 would not be precisely $1/F$ in the present case because  our pions are not in a trivial vacuum,
 but rather in the $\<\pi^-\> $ condensed phase. However, we expect that the
 general algebraic structure of the vacuum expectation value will be obtained correctly using
 this approach. Indeed, taking $ O=   \bar{u} \gamma_0
u   -  \bar{d} \gamma_0 d   $, as the isospin density and
calculating the commutator $[O, Q^5]$, where $Q^5=\int
d^3x\,\bar{u}\gamma_0 \gamma_5 d=\int d^3x\,u^{\dagger} \gamma_5
d$ is the axial charge, one obtains the structure $ \langle i
\bar{u} \gamma_0 \gamma_5 d\rangle $ entering the
eq.\,(\ref{newdensI}). Therefore, if we expect the vacuum
expectation value of $ \<O\>= \<\bar{u} \gamma_0 u   - \bar{d}
\gamma_0 d \> $ to be nonzero, we should also expect a nonzero
value for the axial density $ \langle i \bar{u} \gamma_0 \gamma_5
d\rangle $. This logic is definitely   supported by the explicit
calculations (\ref{newdensI}).

One can also test formula  (\ref{newdensI}) at small isospin
density $n_I$. In this case our system may be understood as a
dilute Bose-Condensate of non-relativistic $\pi^{-}$
particles\cite{kogut2}. How is $n_A$ manifested in this
terminology? We shall work at fixed isospin number density
(instead of at fixed $\mu_I$). Moreover, we will temporarily work
in Minkowski space. In the Isospin phase, the diquarks are not
important as we saw, so we parameterize $\Sigma$ as, \beq \Sigma =
\left(\begin{array}{cc} 0 & -U\\ U^T & 0\end{array}\right), \quad
U \in SU(2)\eeq The field $U$ transforms as $U \to L U
R^{\dagger}$ under $SU(2)_L \times SU(2)_R$ and the effective
Lagrangian for $U$ reads, \beq {\cal L} = F^2 \left( \Tr \d_{\mu}
U \d^{\mu} U^{\dagger} + 2 m^2_{\pi} Re \Tr U\right)\eeq Thus, we
see that the Lagrangian describing the pion sector of $N_c = N_f =
2$ QCD is exactly the same as the one describing $N_c = 3$, $N_f =
2$ QCD. We express, $U = \exp\left(\frac{i \pi^a \sigma^a}{2
F}\right)$. Similarly to eq.\,(\ref{dF}),(\ref{dFeff}), we
identify, \beqn \bar{\psi} \gamma^{\mu} \frac{\sigma^a}{2}\psi &=&
2i F^2 \Tr \left([\d^{\mu} U, U^{\dagger}]
\frac{\sigma^a}{2}\right) \approx
-\epsilon^{abc} \d^{\mu} \pi^b \pi^c\\
\bar{\psi} \gamma^{\mu} \gamma^5 \frac{\sigma^a}{2}\psi &=& -2 i
F^2 \Tr \left(\{\d^{\mu} U, U^{\dagger}\}
\frac{\sigma^a}{2}\right)
\approx 2 F \d^{\mu} \pi^a\\
i \bar{\psi} \gamma^5 \frac{\sigma^a}{2} \psi &=& i g \Tr \left((U
- U^{\dagger})\frac{\sigma^a}{2}\right) \approx -g
\frac{\pi^a}{F}\eeqn where we have expanded the corresponding
currents to leading order in $\pi$ fields. We also expand the
Lagrangian to fourth order in $\pi$ fields, \beq \label{Lquart}
{\cal L} = \frac{1}{2} \d_{\mu} \vec{\pi} \d^{\mu} \vec{\pi} -
\frac{1}{2} m^2_\pi \vec{\pi}^2 - \frac{1}{24 F^2}\vec{\pi}^2
\d_{\mu} \vec{\pi} \d^{\mu} \vec{\pi} + \frac{1}{96
F^2}\d_{\mu}(\vec{\pi}^2) \d^{\mu} (\vec{\pi}^2) + \frac{1}{96
F^2} m^2_{\pi} (\vec{\pi}^2)^2\eeq

We can ignore the $\pi^0$ particles as they are irrelevant for
$\pi^-$ condensation. It is useful to combine $\pi =
\frac{1}{\sqrt{2}}(\pi^1 + i \pi^2)$. To describe non-relativistic
physics involving $\pi^{-}$ particles, we can replace $\d_0 \pi
\to i m_\pi \pi$, $\d_i \pi \to 0$, in the quartic terms of
Lagrangian (\ref{Lquart}). Finally, we adopt a non-relativistic
normalization of our $\pi^-$ field, by introducing, a canonical,
non-relativistic Bose field (of dimension 3/2), \beq
\phi^{\dagger} = \sqrt{2 m_\pi}\, \pi \eeq The Hamiltonian density
in terms of the $\phi$ field reads, \beq H = \frac{1}{2 m_\pi}\d_i
\phi^{\dagger} \d_i \phi + m_\pi \phi^{\dagger} \phi + \frac{1}{32
F^2} (\phi^{\dagger}\phi)^2\eeq while the condensates and
densities become, \beqn n_I &=& \frac{1}{2} \bar{\psi}\sigma_3
\psi = \phi^{\dagger} \phi
\\n_A &=& i \bar{u} \gamma_0 \gamma_5 d = - 2 F \sqrt{m_\pi}
\,\phi^{\dagger}\\i \bar{u} \gamma_5 d &=& - \frac{g}{F
\sqrt{m_\pi}} \,\phi^{\dagger}\eeqn In this language, we see that
both of our $U(1)_I$ order parameters, $n_A$ and $i\bar{u}
\gamma_5 d$ are expressed in terms of the same non-relativistic
Bose field $\phi^{\dagger}$.

The energy density of a spatially uniform Bose-Condensate as a
function of isospin density is, \beq \epsilon = m_\pi n_I +
\frac{1}{32 F^2} n_I^2\eeq Therefore, the isospin chemical
potential, \beq \label{muI}\mu_I = \frac{\d \epsilon}{\d n_I} =
m_\pi + \frac{1}{16 F^2} n_I \eeq One can check that (\ref{muI})
agrees to first order in $n_I$ with the result (\ref{nI}) obtained
in the grand-canonical ensemble treatment. Re-expressing the order
parameters in terms of isospin density, we obtain (up to $U(1)_I$
phase), \beq n_A = 2 F (m_\pi n_I)^{\frac12}, \quad \langle i
\bar{u} \gamma^5 d\rangle = - \frac{1}{2} \pp \left(\frac{n_I}{4
F^2 m_\pi}\right)^{\frac12}\eeq in agreement to leading order with
previous result (\ref{newdensI}), (\ref{pion}).

Thus, the appearance of the second order parameter $n_A$ is quite
natural. Finally, we remark that the situation in the $B$ phase is
the mirror image of the above discussion. The new $U(1)_B$
breaking density is, \beq \langle i u^T \gamma_0 C \gamma_5 \tau_2
d \rangle = -4 F^2 \mu_B \lambda
\left(1-\lambda^2\right)^{\frac12} \cos(\alpha)\eeq

\subsection{$\theta$\, Dependence}

So far we have been mostly investigating the phase diagram in the
$(\mu_B, \mu_I)$ plane at fixed $\theta$. In this section we would
like to focus more on the $\theta$ dependence, drawing the phase
diagram in the $(\theta, \mu)$ plane. This trivial exercise leads
to rather interesting consequences, namely, the $\theta$
dependence at fixed $\mu$ becomes non-analytic. We further
characterize the phase diagram in terms of the $\langle G
\tilde{G}\rangle$ correlator and the topological susceptibility
$\chi$. Finally, we confirm our calculations by checking the
validity of low energy theorems.

To simplify the discussion we shall take $\mu_I = 0$ and focus on
$\theta$ dependence in the Normal and Baryon phases.  The
situation in the Isospin phase is again just the mirror image, as
can be explicitly checked.

We begin by considering $\theta$ dependence at $\mu = 0$. The
story is exactly the same as in the well-studied case $N_c = 3,
N_f = 2$. The vacuum energy density ${\cal F}(\theta)$ is, \beqn
\label{EN} {\cal F}(\theta, \mu = 0) &=& -4 F^2
m^2_\pi(\theta)\eeqn where, \beq m^2_{\pi}(\theta) =
\frac{m(\theta) |\pp|}{4 F^2}\eeq \beq m(\theta) =
\frac{1}{2}\left({(m_u+m_d)^2\cos^2(\theta/2) +
 (m_u-m_d)^2\sin^2(\theta/2)}\right)^{\frac12}\eeq By
differentiating $\cal{F}(\theta)$ we can compute correlation
functions of $G \tilde{G}$, \beqn \frac{\d {\cal F}}{\d \theta}
&=& \langle i \GDG \rangle\\ - \frac{\d^2 {\cal  F}}{\d \theta^2}
&=& \chi = - \int d^4x \langle T \GDG (x) \GDG
(0)\rangle_{conn}\eeqn At $\mu = 0$ we find, \beqn \nn\langle i
\GDG \rangle_{\mu = 0} &=&
-\frac{1}{4}\frac{m_u m_d}{m(\theta)} \, \sin(\theta)\,\pp\\
\chi(\mu = 0) &=& \frac{1}{4}\frac{m_u m_d}{m(\theta)}
\left(\cos(\theta) + \frac{m_u m_d}{4 m(\theta)^2}
\sin^2(\theta)\right) \pp \label{GDG0}\eeqn

The expressions (\ref{GDG0}) reflect the well-known strong
$\theta$ dependence in the region $m_u \approx m_d = m_q$, $\theta
\approx \pi$. Let's introduce the asymmetry parameter, \beq
\epsilon = \frac{|m_u - m_d|}{m_u + m_d}\eeq and assume $\epsilon
\ll 1$.

\begin{figure}[t]
\begin{center}
\includegraphics[angle=-90, width = 0.45\textwidth]{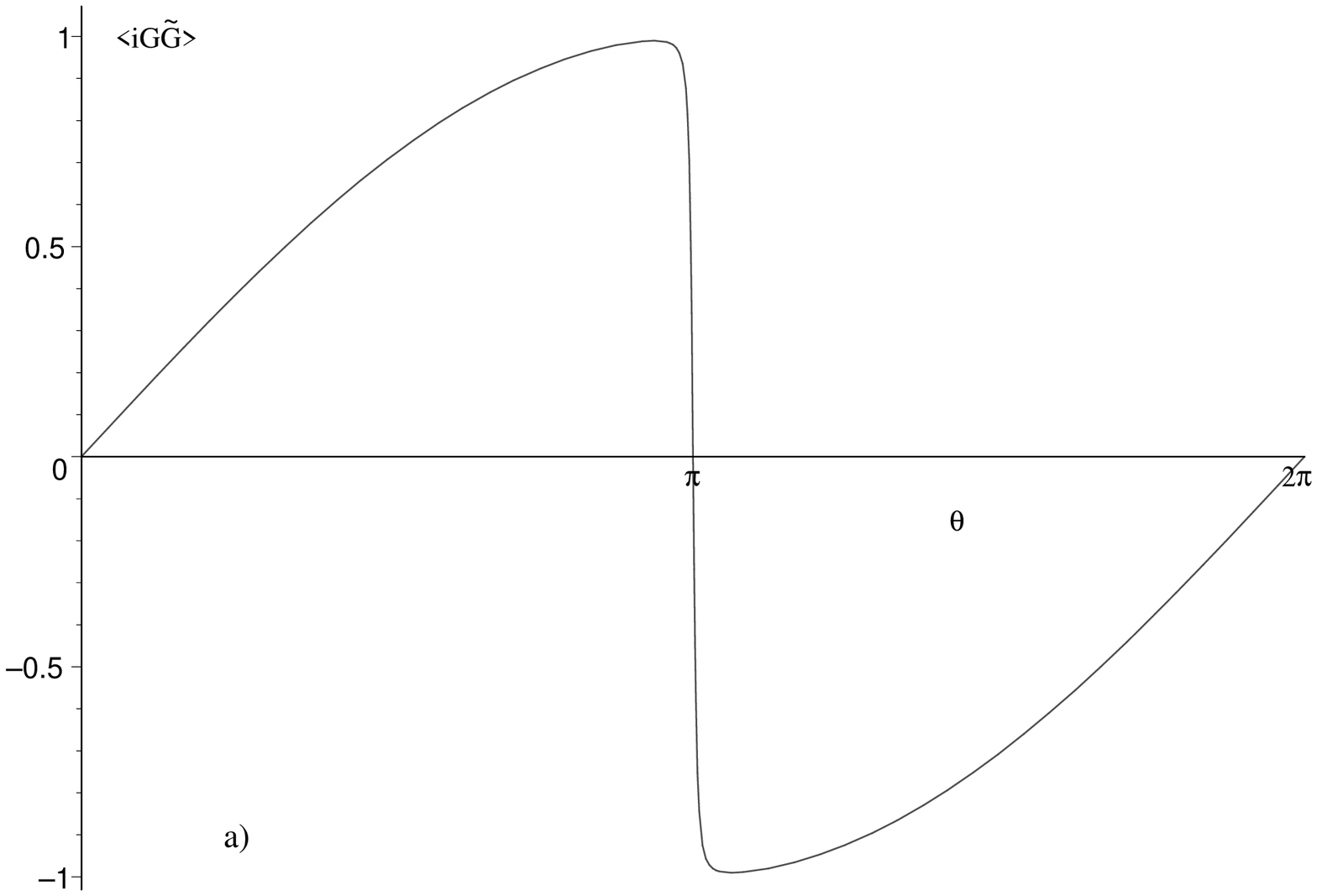}
\includegraphics[angle=-90, width = 0.45\textwidth]{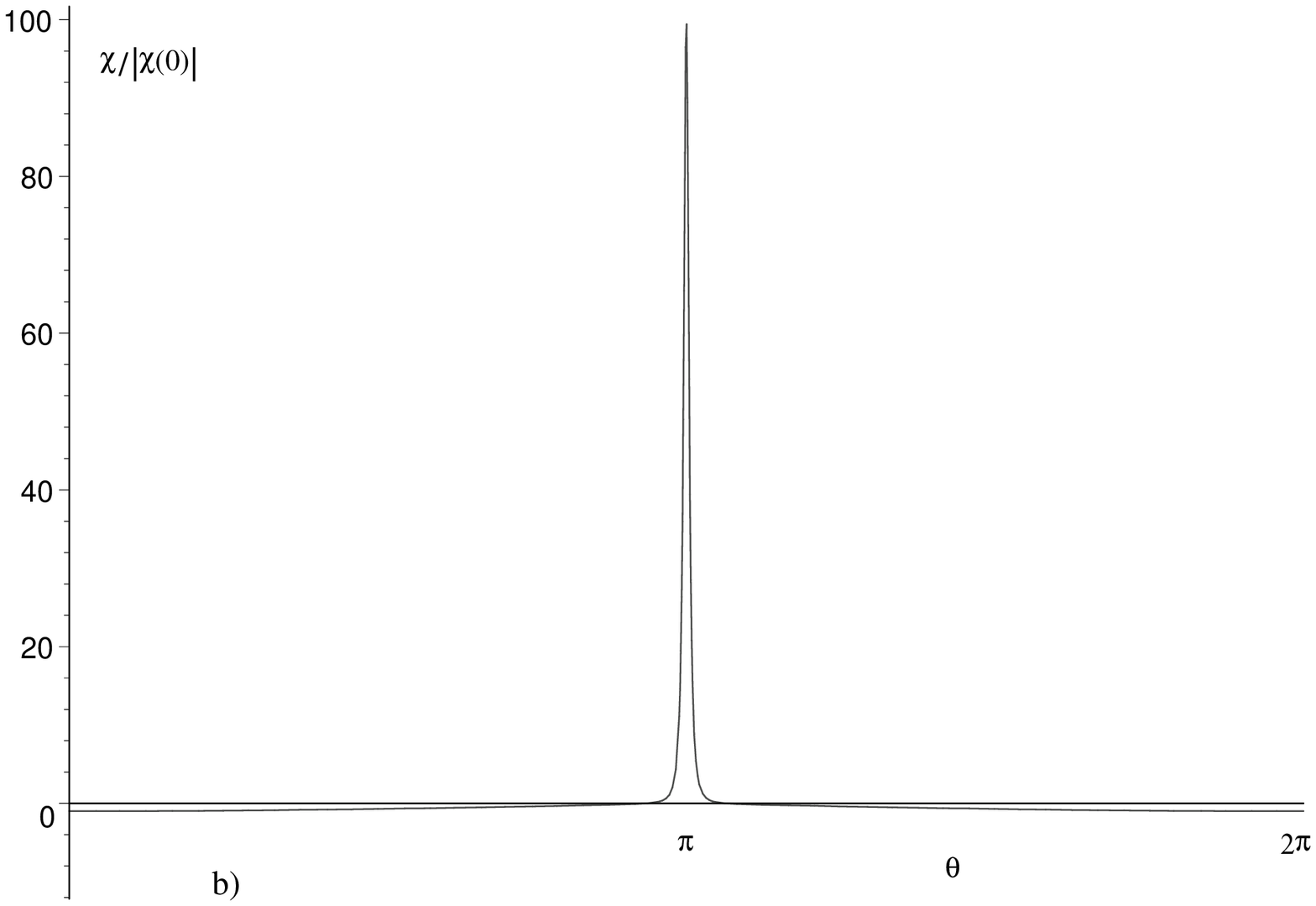}
\caption{$\theta$ dependence in $N_c = 2$, $N_f = 2$ QCD at $\mu =
0$, $\epsilon = 0.01$. a) The $CP$ odd order parameter $\langle i
\GDG\rangle$. See eq.\,(\ref{GDG0}) for precise normalization. b)
Topological susceptibility $\chi$.} \label{chi0}
\end{center}
\end{figure}

The $CP$ odd order parameter $\langle i G \tilde{G}\rangle$ (see
Fig.\,\ref{chi0},a) starts out at $0$ when $\theta = 0$ and
increases smoothly with $\theta$, reaching its maximum just before
$\theta = \pi$ at, \beq \langle i \GDG\rangle_{\theta = \pi^-}
\approx - \frac{m_q}{2} \pp \eeq Afterwards, the order parameter
$\langle i G \tilde{G} \rangle$ experiences a steep crossover,
dropping to its minimum of, \beq \langle i \GDG \rangle_{\theta =
\pi^+} \approx + \frac{m_q}{2} \pp\eeq The crossover occurs in the
region $|\theta - \pi| \sim \epsilon$ and hence, the topological
susceptibility $\chi$ has a sharp peak around $\theta = \pi$ of
width $\Delta \theta \sim \epsilon$ and height,
$\chi(\pi)/|\chi(0)| =  1/\epsilon$ (see Fig.\,\ref{chi0},b).

Such behaviour of the $CP$ odd order parameter $\langle i G
\tilde{G}\rangle$ strongly suggests that for $m_u = m_d$,
spontaneous breaking of $CP$ symmetry occurs at $\theta = \pi$.
This situation, known as Dashen's phenomenon, has been extensively
studied in $N_c = 3$ QCD with  $N_f = 3$ and $N_f =
2$\cite{Dashen,Witten:1980sp,DiVecchia:1980ve,Creutz,Smilga,Tytgat}.
For $N_f = 3$ with $m_s \gg m_u, m_d$ it is believed that
spontaneous $CP$ breaking occurs at $\theta = \pi$ for $|m_u -
m_d| m_s < m_u m_d$. For $N_f = 2$, $CP$ violation occurs at
$\theta = \pi$, $m_u = m_d$ and possibly in a small window of
$|m_u - m_d| \neq 0$\cite{Tytgat}.

However, it is important to note that Dashen's phenomenon is not
under complete theoretical control in our effective
Lagrangian(\ref{Lch}). Indeed, for a moment, we fix $m_u = m_d$.
Then, for general $\theta$, the mass term explicitly breaks the
symmetry of the effective Lagrangian (\ref{Lch}) from $SU(4)$ to
$Sp(4)$. However, for $\theta = \pi$, the mass term in the
effective Lagrangian vanishes, restoring the symmetry to $SU(4)$
and giving rise to apparently massless goldstones:
$m_\pi^2(\theta=\pi) = 0$. Yet, no such symmetry restoration
occurs in the fundamental microscopic QCD Lagrangian at $\theta =
\pi$. This contradiction is resolved by including higher order
(quadratic) mass terms in the effective Lagrangian, which would
explicitly break $SU(4)$ even at $\theta = \pi$\,\cite{Smilga}. It
is precisely these terms, which control the physics of Dashen's
phenomenon, and which are not included in the present work.

We do not wish to consider such higher order mass terms in this
paper. For any fixed $\frac{|m_u - m_d|}{m_u + m_d} \neq 0$ these
terms can be neglected by considering sufficiently small $m_q$. If
the higher order mass terms are largely saturated by a third quark
of mass $m_{u,d} \ll m_s \ll \Lambda_{QCD}$, then we require, \beq
\label{meta}\frac{|m_u - m_d|}{m_u + m_d} \gg
\frac{m_{u,d}}{m_s}\sim \frac{m^2_{\pi}(\theta=0)}{M_\eta^2}\eeq
This condition is, indeed, realized in the true physical world.
If, on the other hand, the higher order terms are controlled by a
light $\eta'$ (as motivated by $N_c \rightarrow \infty$), we
consider, \beq \label{mtau} \frac{|m_u - m_d|}{m_u + m_d} \gg
\frac{m \pp}{F^2_{\pi} M^2_{\eta'}} \sim
\frac{m^2_{\pi}(\theta=0)}{M^2_{\eta'}}\eeq Of course, by imposing
restrictions (\ref{meta}), (\ref{mtau}) we automatically exclude
the regions of parameter space where Dashen's transition is
realized, and we may discuss only the quantitatively steep
crossover in the Normal phase. However, we shall see in a moment
that by considering the system at finite $\mu$, the rapid changes
in the vicinity of $\theta\simeq \pi$ observed in the Normal phase
will be washed out.

\begin{figure}[t]
\begin{center}
\includegraphics[angle=-90, width = 0.7\textwidth]{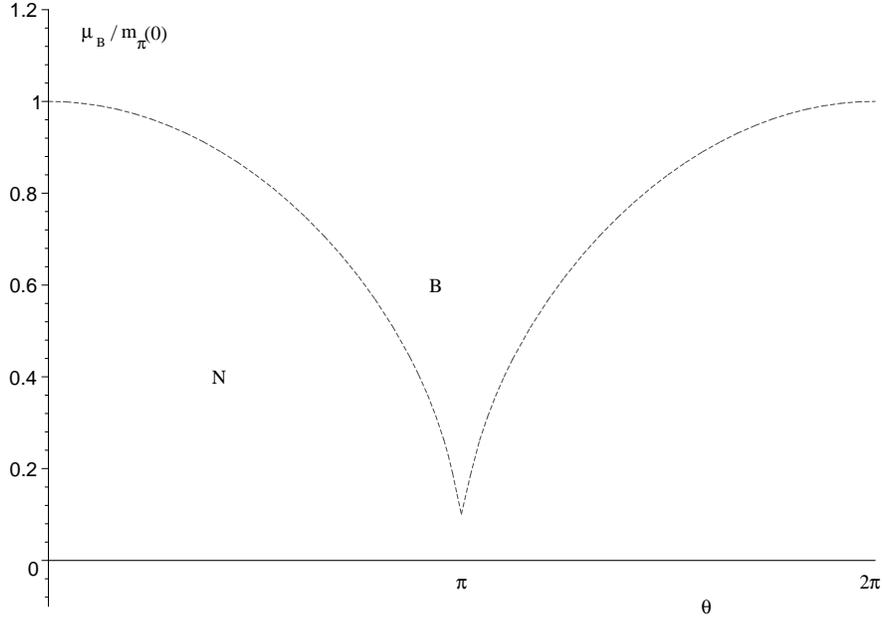}
\caption{Phase diagram of $N_c = 2$, $N_f = 2$ QCD as a function
of $\mu_B$ and $\theta$. Here, $\epsilon = \frac{m_u - m_d}{m_u +
m_d} = 0.01$ A rapid crossover occurs in the Normal phase at
$\theta = \pi$, which is conjectured to become a first order phase
transition, when $m_u = m_d$.}\label{NB}
\end{center}
\end{figure}

Let us now turn on finite $\mu_B$. Once conditions (\ref{meta}),
(\ref{mtau}) are met, all the results of previous sections hold
for any $\theta$. In particular, the transition to the Baryon
phase occurs at $\mu = m_\pi({\theta})$ (see Fig.\,\ref{NB}).
As explained above, we can consider arbitrarily small ratio
$m^2_{\pi}(\theta=\pi)/m^2_{\pi}(\theta=0) = \epsilon$ as long as
$m_q \rightarrow 0$. Thus, for $\mu_B < m_\pi(\theta=\pi)$ the
Normal phase is realized for all $\theta$, while for $\mu_B >
m_\pi(\theta=0)$ we are entirely in the Baryon phase. Finally, if
we fix $\mu_B$ with $m_\pi(\theta=\pi) < \mu_B < m_\pi(\theta=0)$
and vary $\theta$ from $0$ to $2\pi$ we encounter two phase
transitions: from Normal to Baryon phase and then back to Normal.
Thus, the $\theta$ dependence becomes non-analytic in this region!
Since the N to B phase transition is second order, we expect the
topological susceptibility, $\chi$ to be discontinuous across the
phase boundary. The transitions between B and N phases occur at
$\theta = \theta_c$ and $\theta = 2\pi - \theta_c$, with the
critical $\theta_c$ given by $m_\pi(\theta_c) = \mu$.

 In the
Baryon phase, the free energy density reads, \beq \label{EB} {\cal
F}(\theta) = - 2 F^2 \mu^2_B \left(1 +
\frac{m^4_\pi(\theta)}{\mu^4_B}\right)\eeq Clearly, the $\theta$
dependence in the superfluid phase is different from that in the
Normal phase (\ref{EN}). This is most clearly seen by computing,
\beqn \langle i \GDG \rangle &=& \frac{m_u m_d}{16 F^2 \mu_B^2}
{\pp}^2 \sin(\theta),\nn\\ \chi &=& - \frac{m_u m_d}{16 F^2
\mu_B^2} {\pp}^2 \cos(\theta)\label{GDGB}\eeqn We have to remember
that expressions (\ref{GDGB}) hold for all $\theta$ only once we
are entirely in the Baryon phase: $\mu_B > m_{\pi}(\theta=0)$. On
the other hand, if $m_{\pi}(\theta=\pi) < \mu_B <
m_{\pi}(\theta=0)$, then we use expression (\ref{GDG0}), for
$\theta$ where the Normal phase is realized, and expression
(\ref{GDGB}), for $\theta$ where the Baryon phase exists. Focusing
for a moment on $\mu_B > m_{\pi}(\theta=0)$, we see that the
$\theta$ dependence is very smooth: there is no sign of rapid
crossover in $\langle i G \tilde{G}\rangle$ near $\theta = \pi$
and the large peak in the susceptibility $\chi$ disappears.
Moreover, as $\mu_B$ increases, the $\theta$ dependence is
suppressed, as expected. This smooth $\theta $ dependence at
$\theta \sim \pi $ in the superfluid phase should be contrasted
with sharp behavior in the Normal phase discussed above, see Figs
\ref{chi0},a,b.

Now we would like to understand, how the strong $\theta$
dependence at $\mu = 0$ gets smoothed out as the chemical
potential $\mu_B$ increases. For, $0 < \mu_B < m_\pi(\theta=\pi)$,
$\theta$ dependence is the same as at $\mu = 0$. The key region is
$m_{\pi}(\theta=\pi) < \mu_B < m_\pi(\theta=0)$, where at fixed $\mu_B$,
\begin{figure}[h]
\begin{center}
\includegraphics[angle=-90, width = 0.45\textwidth]{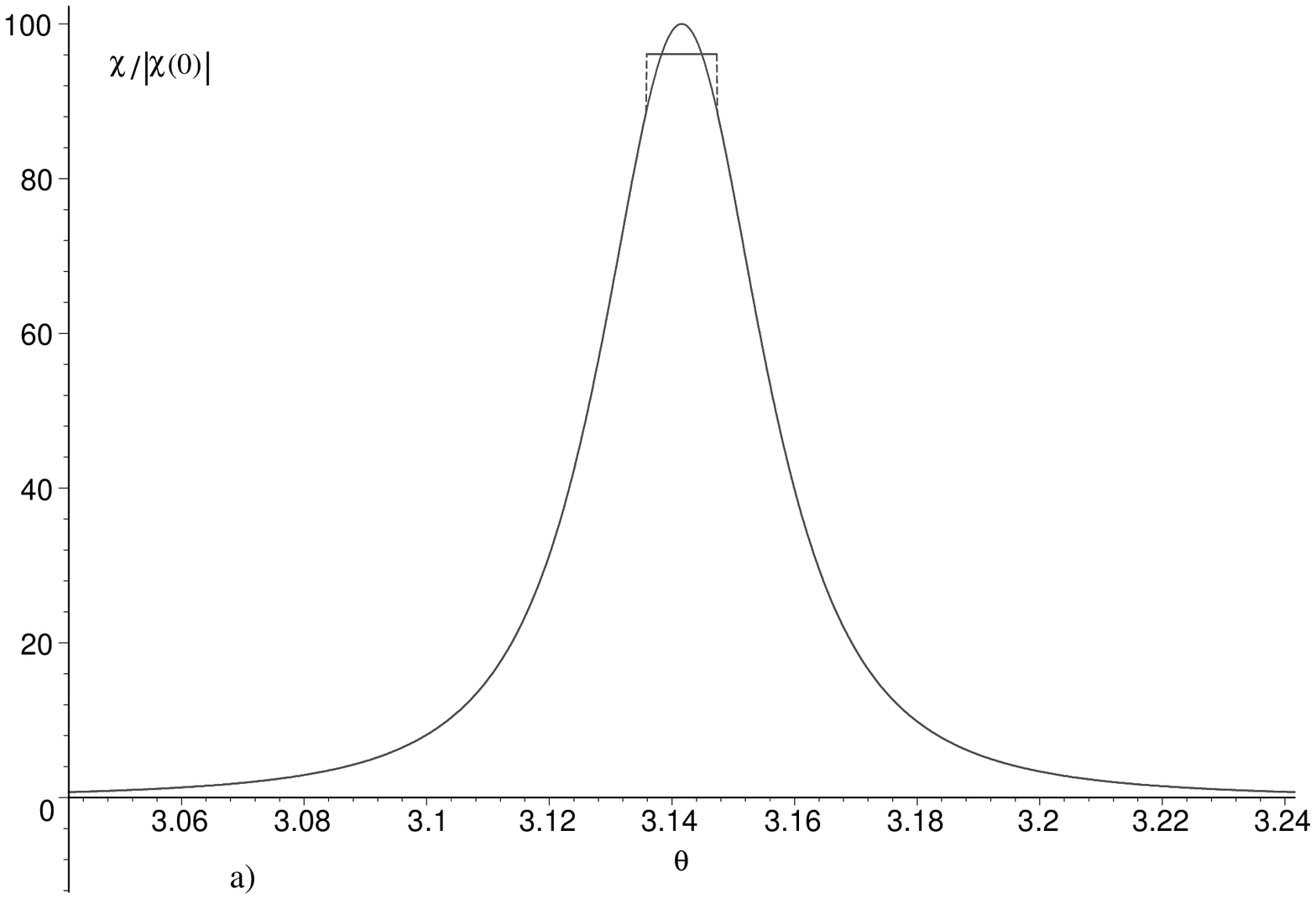}
 \includegraphics[angle=-90, width =
 0.45\textwidth,clip]{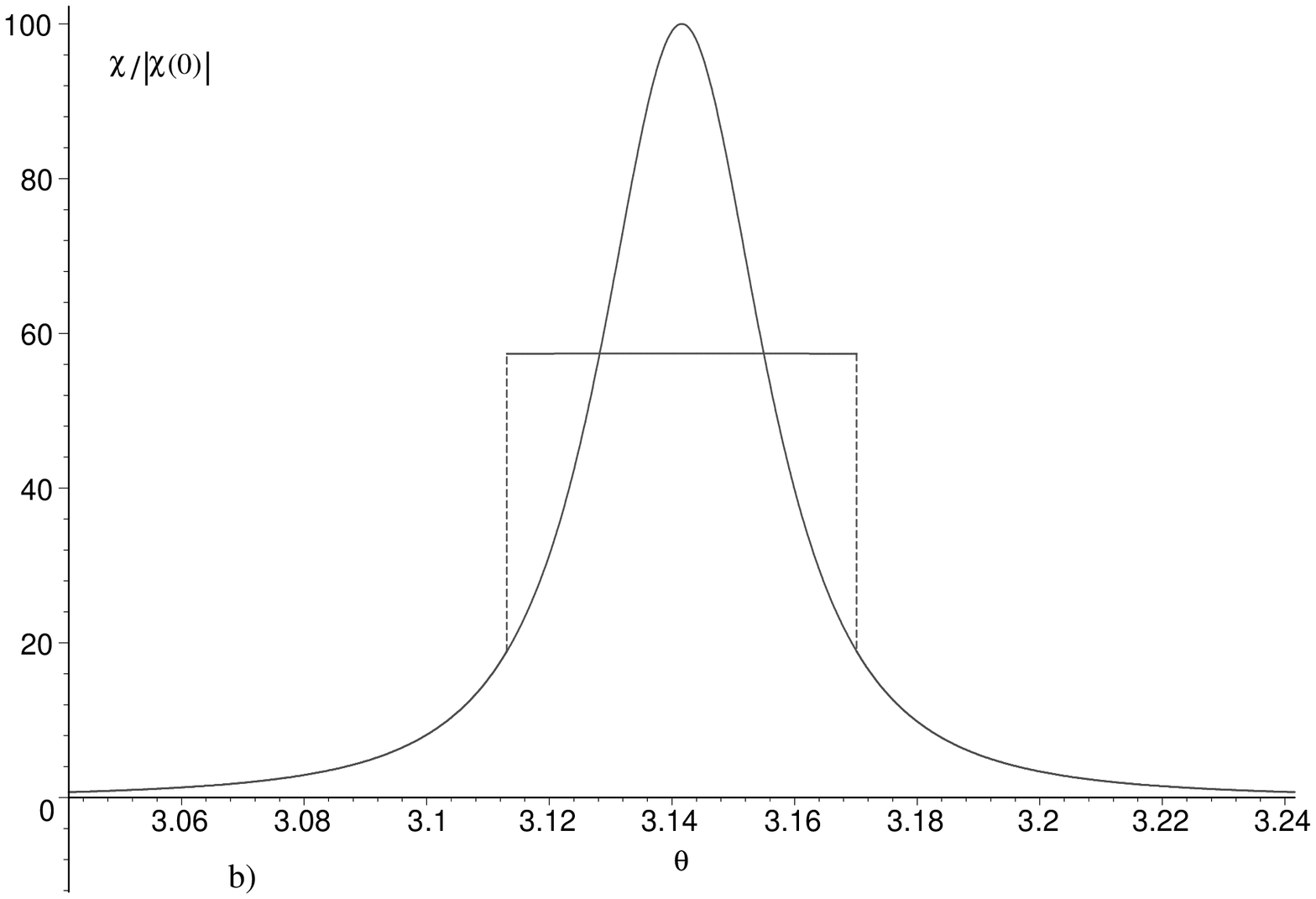}
 \includegraphics[angle=-90, width = 0.45\textwidth,clip]{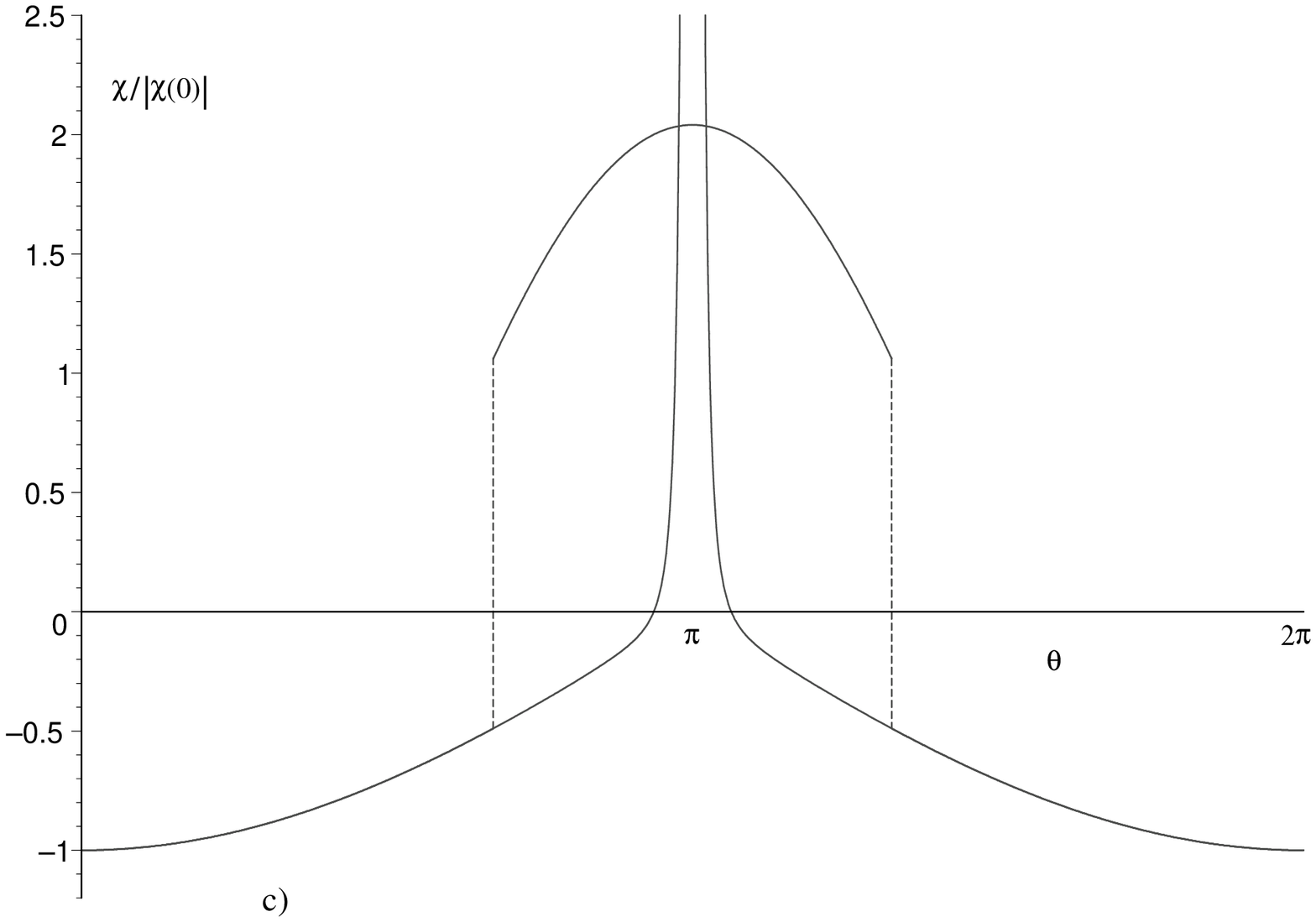}
\includegraphics[angle=-90, width = 0.45\textwidth]{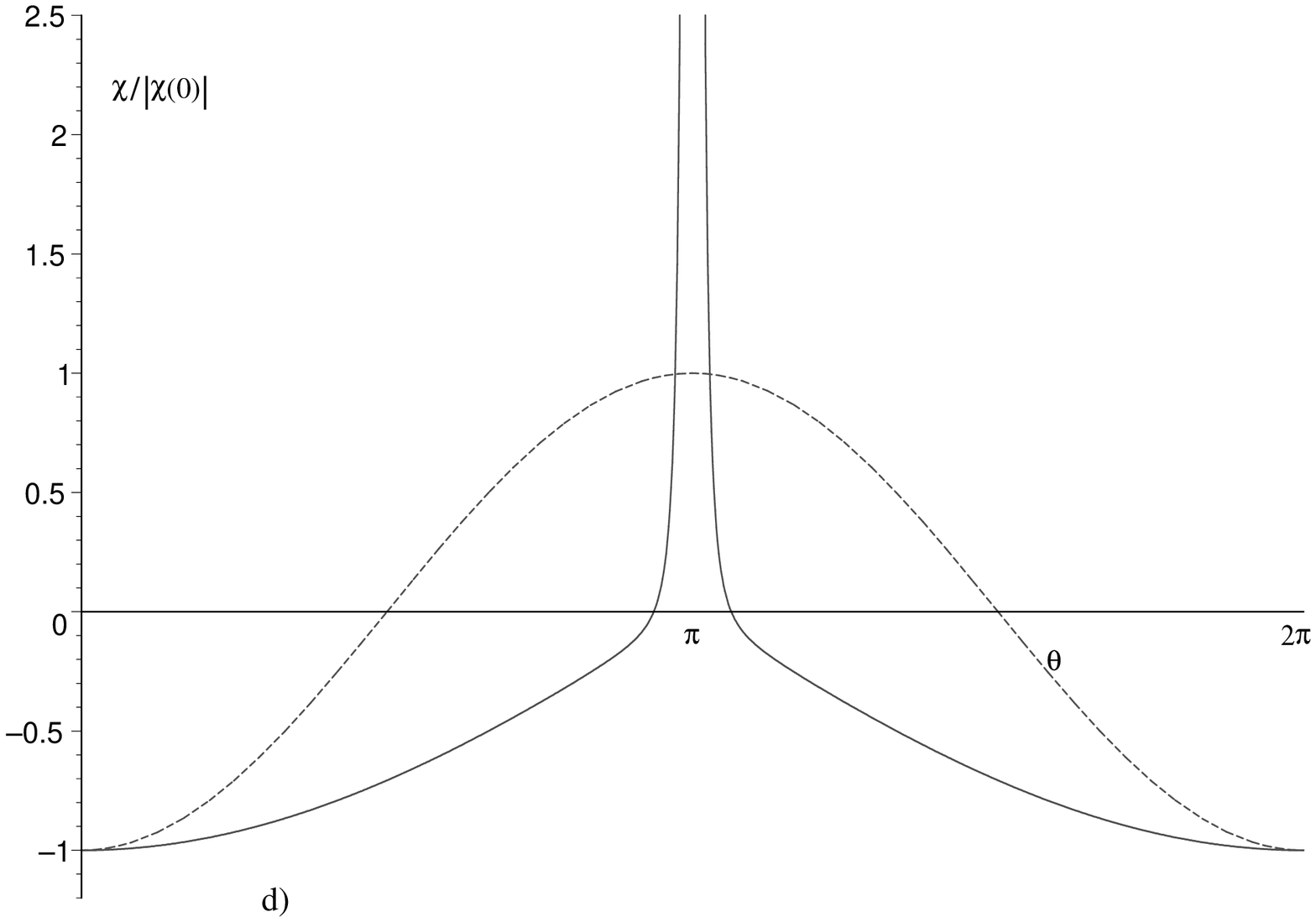}
\caption{a) $\theta$ dependence of the topological susceptibility
$\chi$, in $N_c = 2$, $N_f = 2$ QCD at $\mu = 1.02 \,
m_{\pi}(\theta=\pi)$ (broken curve) together with $\chi$ at $\mu =
0$ (unbroken curve). Here, $\epsilon = 0.01$  b) the same with
$\mu = 1.3 \, m_{\pi}(\theta=\pi)$; c)  the same with $\mu = 0.7
\, m_{\pi}(\theta= 0)$;  d) the same with $\mu =
m_{\pi}(\theta=0)$}\label{chis}
\end{center}
\end{figure}
the Baryon phase is realized for $\theta_c < \theta < 2\pi - \theta_c$
and the Normal phase is realized otherwise.
 Let us investigate the
behaviour of the topological susceptibility $\chi$ in this region.
As we expected, $\chi$ is discontinuous across the phase
transition, \beq \label{jump}\frac{\chi(\theta_c^+) -
\chi(\theta_c^-)}{|\chi(0)|} = \frac{m_u m_d m^2_{\pi}(0)
\pp^2}{64 F^4 \mu^6_B} \sin^2(\theta_c)\eeq As the chemical
potential increases slightly past $m_\pi(\theta=\pi)$, a narrow
region of Baryon phase appears around $\theta = \pi$, deep inside
the crossover region shown on Fig.\,\ref{chi0}. In terms of
susceptibility $\chi$, this affects only the very top of the peak
of $\chi(\theta)$ by introducing small discontinuities at $\theta
= \theta_c$ and $\theta = 2 \pi - \theta_c$ (see
Fig.\,\ref{chis},a). As $\mu$ further increases, the range of
$\theta$ where the B phase exists starts growing. This is
accompanied by growth in discontinuities of $\chi$ at the
transition points. Eventually, for $m_{\pi}(\theta=\pi) \lesssim
\mu \ll m_{\pi}(\theta=0)$, the original peak in $\chi$ associated
with the $CP$ crossover, is entirely replaced by discontinuities
associated with the second order Normal to Baryon phase transition
(see Fig.\,\ref{chis},b). Once this occurs, the magnitude of the
jump $\chi(\theta_c^+) - \chi(\theta_c^-)$ starts to decrease
(Fig.\,\ref{chis},c), until finally at $\mu = m_\pi(\theta=0)$,
$\chi$ becomes continuous again and we are entirely in the Baryon
phase (Fig.\,\ref{chis},d). The washout of the sharp $\theta$
dependence near $\theta = \pi$ has been realized!

\begin{figure}[h]
\begin{center}
\includegraphics[angle=-90, width = 0.7\textwidth]{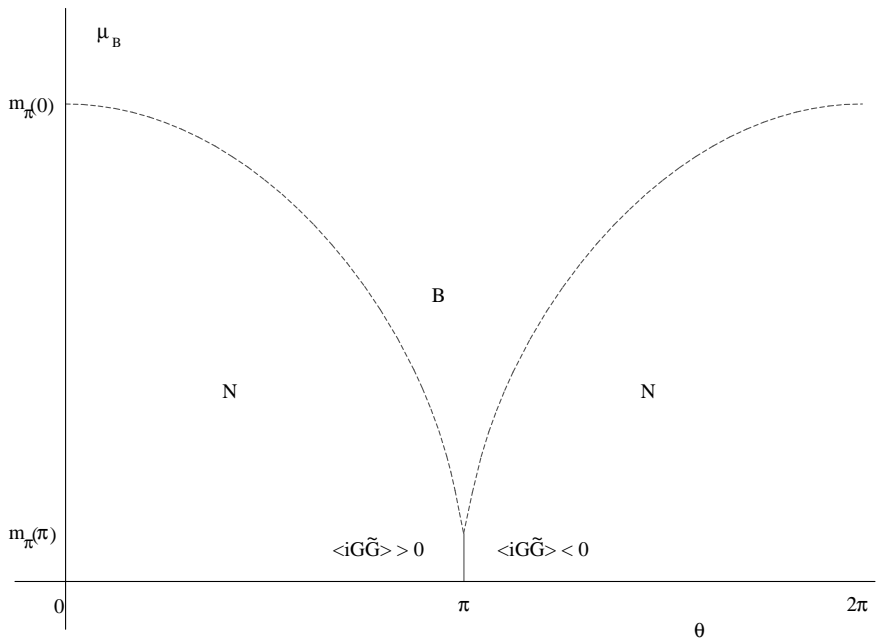}
\caption{Conjectured form of the Phase diagram of $N_c = 2$, $N_f
= 2$ QCD for $m_u = m_d$. Solid line indicates a first order phase
transition, while dashed lines indicate second order phase
transitions. The region near $\theta = \pi$ is not to scale.}
\label{NBN}
\end{center}
\end{figure}

The most exciting result of this section is that for $m_u \neq
m_d$ the ``Dashen's crossover" first splits into two second order
Normal to superfluid phase transitions and for $\mu > m_\pi(0)$
gets entirely washed out. We would like now to provide some
speculations regarding the degenerate case $m_u = m_d$, $\theta =
\pi$. This point might be of importance for lattice
fermions\cite{Creutz,Aoki}, as it is equivalent to a theory where
one quark mass is negative and $\theta$ parameter is not
explicitly present. In principle, it is possible to analyze this
situation rigourously by going to higher order Chiral Perturbation
Theory. However, since the algebra becomes rather involved even at
$\mu = 0$, we confine ourselves to a conjecture based on the above
results and common $\theta = \pi$ lore. For $N_c = 2$, $N_f = 2$,
at $\mu = 0$, we expect that Dashen's phenomenon will occur along
the same lines\cite{Smilga} as for $N_c = 3$. Spontaneous $CP$
violation will happen at $\theta = \pi$, however, no continuous
symmetries will be broken and Goldstones will have a small, but
finite mass $m_\pi(\theta = \pi) > 0$. At finite $\mu$, we expect
a line of first order phase transitions at $\theta = \pi$ to
extend to $\mu = m_\pi(\theta = \pi)$, where it splits into two
second order phase transition lines (see Fig.\,\ref{NBN}). We
remark that in the Baryon phase, P-parity is still spontaneously
broken at $\theta = \pi$, while in the Isospin phase, P-parity is
broken at $\theta = 0$, but not a $\theta = \pi$.

\section{Ward Identities}
In  this section we check the validity of Ward Identities (WI)
\cite{DiVecchia:1980ve, WI, Witteneta, NSVZ} at nonzero $\mu$,
$\theta$. We anticipate that WI must remain untouched when
external parameters such as $\mu$, $\theta $  or temperature $T$
are introduced.  Indeed, the anomaly (chiral and/or conformal) is
a short distance ($UV$) phenomenon, which is not affected by
medium effects (density $\mu \neq 0$, $\theta$ and/or temperature
T). This fact was implicitly used when we constructed the
effective Lagrangian (\ref{Lch}). However, we are in a position to
calculate each term entering  the WI explicitly. Therefore, the
check of the WI is a nontrivial test of self consistency of our
results.

The first identity that we consider, relates the two $CP$ odd
order parameters, \beq \label{WI1} \langle i \GDG \rangle =
\frac{1}{N_f} \langle i \bar{\psi} \gamma_5 M \psi\rangle\eeq The
identity (\ref{WI1}) reflects the well known fact that there is no
$\theta$ dependence when $m \rightarrow 0$. By consulting Table 1
and eqs. (\ref{GDG0}), (\ref{GDGB}), we can explicitly check that
our results satisfy the identity (\ref{WI1}) both in Normal and
superfluid phases.

The next WI we would like to discuss is, \beq \label{WI2p}\chi = -
\int d^4x \langle T \GDG(x) \, \GDG(0)\rangle_{conn} =
\frac{1}{N^2_f}\langle \bar{\psi} M \psi\rangle + O(M^2)\eeq The
$O(M^2)$ term in (\ref{WI2p}) is usually dropped in the chiral
limit, $SU(N_f)_V$ symmetric limit at $\theta=0$, assuming  the
resolution of the $U(1)$ problem when flavor singlet $\eta'$ is a
heavy state. Indeed, in this case Table 1 and
eqs.\,(\ref{GDG0}),(\ref{GDGB}) imply that the WI (\ref{WI2p})
holds both in Normal and superfluid phases. An important remark is
that both sides of (\ref{WI1}) and (\ref{WI2p}) depend on $\mu$ in
a very nontrivial way. Nevertheless, the identities are preserved
as expected.

Now, we would like to see what happens with (\ref{WI2p}) when we
relax the requirement of the $SU(N_f)_V$ symmetric limit and also
consider $\theta \neq 0$. In this case  it is important to keep
the $O(M^2)$ term,
 \beq O(M^2) = -
\frac{1}{N^2_f} \int d^4x \langle T \bar{\psi} \gamma_5 M \psi(x)
\, \bar{\psi} \gamma_5 M \psi(0)\rangle_{conn}\eeq

We begin in the Normal phase and evaluate, \beq \chi -
\frac{1}{N^2_f} \langle \bar{\psi} M \psi \rangle = - \frac{1}{64}
\frac{(m^2_u - m^2_d)^2}{m(\theta)^3} \pp \eeq The above result
implies that $m_q^{-1}$ singularities develop in the $O(M^2)$ term
of eq.\,(\ref{WI2p}), due to $\eta'$/goldstone mixing, which
occurs for $m_u\neq m_d$.
 However, the singularities disappear
when $m_u = m_d$, so that the $O(M^2)$ term can be neglected in
the chiral, $SU(N_f)_V$ limit, as long as we are sufficiently far
from $\theta = \pi$. This is the physically expected result.

What happens with (\ref{WI2p}) in the superfluid phase, when
$\theta \neq 0$? We can immediately see that the $O(M^2)$ term can
no longer be neglected even when $m_u = m_d$. As has been
discussed above, the Normal to superfluid transition is second
order, so that the topological susceptibility $\chi$ generically
experiences a jump across the phase boundary (\ref{jump}), while
the chiral condensate $\langle \bar{\psi} M \psi \rangle$ (Table
1) is continuous. Thus, for the WI (\ref{WI2p}) to hold, the
$O(M^2)$ term must jump across the phase boundary, accounting for
the discontinuity in $\chi$ and, thus, contributing on the same
footing as $\langle \bar{\psi} M \psi\rangle$ to the righthand
side of (\ref{WI2p}). It is not surprising that the $O(M^2)$ term
becomes important. Indeed, from Table 1 and (\ref{GDGB}) we see
that both $\chi$ and $\langle \bar{\psi} M \psi\rangle$ are of
order $M^2/\mu^2_B$ in the superfluid phase. The fact that $\chi$
becomes $O(M^2)$ rather than $O(M)$ is part of smoothing out of
$\theta$ dependence in the superfluid phase. The contact between
$O(M)$ dependence in the Normal phase and $O(M^2)$ dependence in
the superfluid phase is provided by the fact that $\mu^2 \sim
O(M)$ at the Normal to superfluid phase transition. Thus, all
correlators in eq.\,(\ref{WI2p}) develop $\mu^{-2}$ singularities
in the superfluid phase, which are due to the modification of the
goldstone spectrum (\ref{SB}).

Finally, from (\ref{WI2p}), to leading order in $M$, in the
superfluid phase, \beq \int d^4x \langle T \bar{\psi} \gamma_5 M
\psi(x) \, \bar{\psi} \gamma_5 M \psi(0)\rangle_{conn} =
-\frac{1}{16 F^2 \mu_B^2} (m^2_u + m_d^2 - 2 m_u m_d \cos(\theta))
\pp^2\eeq which vanishes only if $m_u = m_d$, $\theta = 0$.

\section{Gluon Condensate}
Having determined the $\theta$ and $\mu$ dependence of different
condensates and densities containing
 the quark degrees of freedom (Tables 1, 2),
 one can wonder if similar results  can be derived  for the gluon condensate $\<
 G_{\mu\nu}^2\>$,
 which  describes the gluon degrees of freedom. As is known, the gluon condensate represents
 the vacuum energy of the ground state in the limit $m_q=0,\,\mu=0$ and plays a crucial
 role in such models as the MIT Bag model, where a phenomenological ``bag constant" $B$
describes  the non- perturbative  vacuum energy of the system. The
question we want to answer: how will the gluon condensate $\<
G_{\mu\nu}^2\>$ (bag constant $B$) depend on $\mu, \theta$ if the
system is placed into dense matter? This question is relevant for
a number of different studies such as the equation of state in the
interior
 of neutron stars, see e.g.\cite{Alford:2004pf}, or stability of dense strangelets\cite{Zhitnitsky:2002qa}.
Of course, it is difficult to answer this question in full $3$
color QCD at finite $\mu_B$, however, the answer can be easily
obtained in $2$ color QCD for $\mu \ll \Lambda_{QCD}$, which is
the subject of the present work. We limit ourselves to considering
only the Normal and Baryon phases, the results in the Isospin
phase, as always, are obtained by replacing $\mu_B
\rightarrow\mu_I$.
We work in Minkowski space in this section.

We start from the equation for the conformal anomaly, \beq
\label{Tmm} \Theta^{\mu}_{\mu} = - \frac{b g^2}{32 \pi^2}
G^a_{\mu\nu} G^{a\mu \nu} + \bar{\psi} M \psi \eeq where we have
taken the standard $1$ loop expression for the $\beta$ function
and $b = \frac{11}{3} N_c - \frac{2}{3} N_f = 6$ for $N_c = N_f =
2$. As ususal, a perturbative constant is subtracted in expression
(\ref{Tmm}). For massless quarks and in the absence of chemical
potential, eq.\,(\ref{Tmm}) implies that the QCD vacuum carries a
negative non-perturbative vacuum energy due to the gluon
condensate.

Now, we can use the effective Lagrangian (\ref{Lch}) to calculate
the change in the trace of the energy-momentum tensor $\langle
\theta^{\mu}_{\mu}\rangle$ due to a finite chemical potential
$\mu_B \ll \Lambda_{QCD}$. The energy density $\epsilon$ and
pressure $p$ are obtained from the free
energy density ${\cal F}$, \beqn \epsilon &=& {\cal F} + \mu_B n_B\\
p &=& -{\cal F} \eeqn Therefore, the conformal anomaly implies,
\beq \label{DG2}\langle\cond\rangle_{\mu, m, \theta} -
\langle\cond\rangle_0 = -4 \left({\cal F}(\mu, m, \theta) - {\cal
F}_0\right)- \mu_B n_B(\mu, m, \theta) + \langle \bar{\psi} M \psi
\rangle_{\mu, m,\theta}\eeq Here, the subscript $0$ on an
expectation value means that it is evaluated at $\mu = m = 0, \,
\theta = 0$. The good news is that we have already calculated all
quantities on the righthand side of eq.\,(\ref{DG2}) - see
expressions (\ref{EN}), (\ref{EB}) and Tables 1, 2. Thus, in the
Normal phase we obtain, \beq \label{G2N}\langle\cond\rangle_{\mu,
m,\theta} - \langle \cond\rangle_0 = - 3 m(\theta) \pp \eeq When
$\theta = 0$, (\ref{G2N}) reduces to the standard
result\cite{NSVZ}, which was derived in a different manner. As
expected, $\langle G^2_{\mu\nu}\rangle$ does not depend on $\mu$
in the Normal phase. The Baryon phase is more exciting, \beq
\label{G2B} \langle\cond\rangle_{\mu, m, \theta} -
\langle\cond\rangle_0 = 4 F^2 \mu^2_B \left(1 + 2
\frac{m^4_{\pi}(\theta)}{\mu^4_B}\right).\eeq It is instructive to
represent the same formula in a somewhat different way,
 \beq
\label{G2M} \langle\cond\rangle_{\mu, m, \theta} -
\langle\cond\rangle_{\mu=0, m, \theta} = 4 F^2 (\mu^2_B
-m^2_{\pi}(\theta)) \left(1 - 2
\frac{m^2_{\pi}(\theta)}{\mu^2_B}\right),\eeq which makes contact
with the fact that in the Normal phase, when $\mu_B\leq
m_{\pi}(\theta)$, the gluon condensate does not vary with $\mu_B$.
However, for $\mu_B\geq m_{\pi}(\theta)$, the dependence of the
gluon condensate $\langle G^2_{\mu \nu}\rangle$ on $\mu_B$ in the
Baryon phase becomes  rather interesting. The condensate decreases
with $\mu_B$ for $m_\pi < \mu_B < 2^{1/4} m_\pi$ and increases
afterwards. The qualitative difference in the behaviour of the
gluon condensate for $\mu_B \approx m_\pi$ and for $m_\pi \ll
\mu_B \ll \Lambda_{QCD}$ can be explained as follows. Right after
the Normal to Baryon phase transition occurs, the baryon density
$n_B$ is small and our system can be understood as a weakly
interacting gas of diquarks. The pressure of such a gas is
negligible compared to the energy density, which comes mostly from
diquark rest mass. Thus, $\langle\Theta^{\mu}_{\mu}\rangle$
increases with $n_B$ and, according to the anomaly equation
(\ref{Tmm}), $\langle G^2_{\mu\nu}\rangle$ decreases. A similar
decrease in $\langle G^2_{\mu\nu}\rangle$ with baryon density is
expected to occur in ``dilute" nuclear matter (see \cite{G2orig}
and review \cite{G2rev}). On the other hand, for $\mu_B \gg
m_{\pi}$, energy density is approximately equal to pressure, and
both are mostly due to self-interactions of the diquark
condensate. Luckily, the effective Chiral Lagrangian (\ref{Lch})
gives us control over these self-interactions as long as $\mu_B
\ll \Lambda_{QCD}$. Such control is largely absent in
corresponding calculations of $\langle G^2_{\mu \nu}\rangle$ in
nuclear matter. As $\Delta \epsilon \sim \Delta p$, the trace
$\langle\Theta^{\mu}_{\mu}\rangle$ decreases and the gluon
condensate increases with baryon density. Such behaviour of
$\langle G^2_{\mu \nu}\rangle$ is quite unusual, as finite baryon
density, on general grounds, is expected to suppress the gluons.

At small baryon density, we can also use a slight variant of the
above method for calculation of the gluon condensate, originally
developed in the context of nuclear matter. As long as our
Bose-condensate of diquarks is dilute, we may neglect interactions
between diquarks, and approximate the change in $\langle
G^2_{\mu\nu}\rangle$ as just the expectation value of
$G^2_{\mu\nu}$ in each diquark times their number, \beq
\label{G2q}\langle\cond\rangle_{\mu, m,\theta} -
\langle\cond\rangle_{0, m,\theta} = \frac{n_B}{2 m_\pi} \langle
q^-| \cond |q^-\rangle \eeq Here $q^-$ denotes a diquark state
relativistically normalized to $\langle q^-(p)|q^-(p')\rangle = 2
E_p (2 \pi)^3 \delta^3(p-p')$, giving rise to the factor
$\frac{1}{2 m_\pi}$ in (\ref{G2q}). It remains to calculate the
matrix element, $\langle q^-|G^2_{\mu\nu}|q^-\rangle$. This can be
done by sandwiching the anomaly equation (\ref{Tmm}) between two
diquark states. As $\langle q^-| \Theta^{\mu}_{\mu}|q^- \rangle =
2 m^2_\pi$, \beq \label{G22} 2 m^2_\pi = -\langle
q^-|\cond|q^-\rangle + \langle q^-| \bar{\psi} M
\psi|q^-\rangle\eeq We are used to the fact that the goldstone
mass comes entirely from the symmetry breaking term $\bar{\psi} M
\psi$, so we might, naively, expect from eq.\,(\ref{G22}) that
$G_{\mu\nu}^2$ vanishes in a diquark. However, the diquark mass,
to first order in $M$ is given by, \beq m_\pi^2 = \langle
q^-|\bar{\psi} M \psi|q^-\rangle\eeq therefore, \beq \langle
q^-|\cond|q^-\rangle = -m^2_\pi\eeq so that $G^2_{\mu\nu}$ and
$\bar{\psi} M \psi$ contribute equally to the goldstone mass in
eq.\,(\ref{G22}). Now from eq.\,(\ref{G2q}), \beq
\langle\cond\rangle_{\mu, m,\theta} - \langle\cond\rangle_{0,
m,\theta} = -\frac{1}{2} n_B m_\pi\eeq This is in agreement with
our full result (\ref{G2B}),(\ref{G2M}) to leading order in $n_B$.

Finally, we note that by differentiating (\ref{G2N}), (\ref{G2B})
with respect to $\theta$ we can obtain correlation functions of
$G^2_{\mu\nu}$ with $G \tilde{G}$, in Normal and superfluid
phases.

\section{Conclusion.  Speculations.}
We conclude this paper with the following speculative remarks:\\
\\
{\bf 1.} Naively, one could say that we studied in the present
paper a purely academic question by considering $N_c=2$ rather
than the realized in nature QCD with $N_c=3$.
 We should comment on this
as follows.  First, for $N_c=2$, the so-called, diquarks become
well-defined gauge invariant objects. However, diquarks, as has
been argued  in a number of papers, (see, e.g. recent papers on
the subject\cite{W,WW,Shuryak:2003zi,Exotica,SV,Jaffe}) may play an
important role in 3-color
 QCD dynamics. If the passage from $SU(2)_{color}$ to $SU(3)_{color}$
 does not  lead to dramatic  disturbances of these diquarks,
  these predictions based on $SU(2)_{color}$
 remain qualitatively valid in real QCD!
Arguments supporting the conjecture of smoothness of the
transition $SU(2)_{color}$ to $SU(3)_{color}$ are presented in
\cite{SV}. We also note that there is some similarity between the
proposal of \cite{Jaffe} and the present work to study the diquark
dynamics. In the proposal \cite{Jaffe} the idea is to introduce a
color source to study the diquark dynamics, while in our paper
with $N_c=2$ the diquarks automatically become gauge invariant
objects, and no source is required to study them.

{\bf 2.} It has been suggested that the $\theta$-induced  $CP$ odd
state can be formed in heavy ion collisions  at RHIC, see original
papers\cite{CP} and a recent review\cite{Kharzeev}. Our analysis
could be quite relevant for the study of the decays of a $CP$ odd
configuration, if it is formed.

{\bf 3.} It has been known for quite some time that violation of
parity and CP parity (which is the case when $\theta\neq 0$) may
completely change the phase structure of a theory. Some lattice
calculations, for example, suggest that the behavior of the system
could be very nontrivial when $\theta\neq 0$, and some singular
behavior and even phase transitions are expected, see
e.g.\cite{lattice}. In an environment where $C$, $P$ and $CP$ are
strongly violated, the interaction of quarks and anti-quarks is
not identical, as it is usually assumed, but rather, could be very
different. Under such conditions the QCD phase transition in the
early universe could have a much more complicated history than it
is typically assumed. In particular, one can imagine that some
very nontrivial objects, such as Witten's nuggets\cite{Witten},
which behave as dark matter particles, can be formed. Moreover,
due to the differences in interactions between quarks and
anti-quarks in the presence of $\theta$, local separation of
baryon charges may take place, and
     chunks of quarks or
anti-quarks in condensed color superconducting phase may form
during the QCD phase transition, serving as dark matter\cite{OZ}.
This scenario is based on the idea that while the universe is
globally symmetric, the anti-baryon charge can be stored in chunks
of dense color superconducting antimatter.
  In this case, instead of baryogenesis,
one should discuss the separation of baryon charges.
Such a global picture of our universe, is definitely,  not in contradiction with  the present
observational constraints\cite{OZ}. Rather, it may give a natural explanation for some
global parameters such as $\Omega_{DM}/\Omega_B$\cite{OZ}, or even can naturally explain
the famous $511 $KeV line from the bulge of our galaxy\cite{OZ1}.

Typical relaxation time for $\theta$ is much larger than
$\Lambda_{QCD}^{-1}$, therefore, one can neglect the dynamics of
$\theta$ for studying  the possible phases for each given
$\theta$.  This was the main reason for us to study $\theta$
dependence of the $QCD$ phase diagram. We find in the present
analysis that, indeed, two phase transitions of the second order
will take place when $\theta$ relaxes from $\theta=2\pi $ to
$\theta=0$. These phase transitions will occur under very generic
conditions when $\mu_I$ is smaller than $m_\pi$, but of the
 same order of magnitude as $m_{\pi}$.
The physical consequences of these phase transitions are still to be explored.

 {\bf 4.} Aside from these far reaching speculations, we would like
to mention here a few much more terrestrial consequences of the
present study, which may have some
 impact on the lattice simulations. First of all, 2 color QCD is a nice laboratory to study
 a variety of different very deep problems of gauge theories at nonzero temperature and density, see e.g.
 \cite{Hands,kogut,Kogut:2001if,Akemann:2004qt}.
 New elements, which were not studied previously and which are the subject of the present work,
are related to $\theta$ dependence of different condensates.
There are a few interesting observations which deserve to be mentioned here:\\
a) At $\theta=\pi$, when the determinant of the Dirac operator is
real, the superfluid phase is realized at much lower critical
chemical potential $\mu_c$ than at $\theta=0$. In the limit
$m_u=m_d$, we expect, $\frac{\mu_c(\theta = \pi)}{\mu_c(\theta =
0)} \sim \left(\frac{m}{\Lambda_{QCD}}\right)^\frac12$.
 It gives a unique chance to study the superfluid phase on the lattice at a much smaller $\mu$ than would normally
 be required. We hope that our conjecture for the disappearance of Dashen's transition in the superfluid phase can be explicitly tested on the lattice.\\
 b) Knowledge of $\theta$ dependence of different condensates allows one to calculate the topological susceptibility
  and other  interesting correlation functions as a function of $\mu$. Corresponding  Ward Identities at nonzero $\mu$
can be tested on the lattice.\\
c)  Physics of gluon degrees of freedom and $\mu$ dependence of
the gluon condensate can also be tested on the lattice. The
behavior of the gluon condensate as a function of $\mu$ is very
nontrivial, as has been explained in the text. Nevertheless, our
prediction is robust in a sense that it is based exclusively on
the chiral dynamics and no additional assumptions have been made
to derive the corresponding expression.

Finally, we should emphasize  that all results presented above are
valid only for very small chemical potentials $\mu_B,\mu_I\ll
\Lambda_{QCD}$ when the chiral effective theory is justified. For
larger chemical potentials  we expect a transition to a deconfined
phase at $\mu_B (\mu_I) \simeq 7\Lambda_{QCD}$ \cite{TZ}. We
should also add that all results presented above can be easily
generalized to $N_c=3$ QCD with $\mu_I\neq 0$, $\mu_B=0$\cite{MZ}.

 \section*{Acknowledgements}
  This work
 was supported, in part, by the Natural Sciences and Engineering
Research Council of Canada.

\appendix
\section{Parametrization of the Vacuum Manifold}
The purpose of this appendix is to clarify some global aspects of
the manifold of goldstone fields in $N_c = 2,\, N_f = 2$ $QCD$.
Once the vacuum manifold is parameterized, we show that to first
order in quark mass, all effects of the $\theta$ parameter can be
incorporated into a common real quark mass.

We begin with the assumption that the chiral symmetry breaking
pattern of $N_c = 2,\, N_f = 2$ $QCD$ is, exactly, $SU(4)
\rightarrow Sp(4)$, so that the vacuum manifold ${\cal X} =
SU(4)/Sp(4)$. We represent the manifold as, \beq {\cal X} = \{U
\Sigma_c U^T,\, U \in SU(4)\}\eeq with $\Sigma_c$ given by
eq.\,(\ref{Sigmac}). Note that ${\cal X} \subset {\cal A}$, where
${\cal A}$ is the set of all $4\times4$ antisymmetric, unitary
matrices of determinant $1$. The original work\cite{kogut2} had
implicitly assumed that ${\cal X} = {\cal A}$. As we shall show,
this is almost, but not quite true. In fact, ${\cal A} = {\cal X}
\,\dot\cup \,i {\cal X}$, i.e ${\cal A}$ is a disjoint union of
two pieces, both of which are homeomorphic to $SU(4)/Sp(4)$.

Even though such technical details do not affect the analysis of
\cite{kogut2}, they become important, once the $\theta-$parameter
is introduced. In particular, if we minimized the effective
Lagrangian (\ref{Lch}) over $\Sigma \in {\cal A}$, we would obtain
very different $\theta$ dependence. In fact, the theory
(\ref{Lch}) with $\Sigma\in i{\cal X}$ corresponds to the theory
with $\Sigma\in{\cal X}$ with the redefinition $\theta \rightarrow
\theta + \pi$. As long as we represent our vacuum manifold by any
one of the two pieces ${\cal X}$ or $i {\cal X}$, we obtain the
same physics, if we define $\theta$ appropriately. However, if we
enlarge the vacuum manifold to contain both pieces, the physics
changes: we obtain cusps in $\theta$ dependence at $\theta = \pm
\frac{\pi}{2}$, instead of Dashen's phenomenon at $\theta = \pi$.
Since we find no evidence of an additional spontaneously broken
discrete symmetry that would connect the two disjoint pieces of
${\cal A}$, we insist on our original assumption that the vacuum
manifold is $SU(4)/Sp(4) = {\cal X}$.

Now, let us demonstrate the above claims. We begin by writing any
$\Sigma \in {\cal A}$ as, \beq \label{form}\Sigma =
\left(\begin{array}{cc} a\, \sigma_2 & C\\ -C^T & b
\,\sigma_2\end{array}\right)\eeq Here, a, b are complex numbers
and we have used the fact that $\Sigma$ is antisymmetric. The
requirement, $\det\Sigma = 1$, implies, \beq (det(C) + a b)^2 =
1\eeq  We call, ${\cal A}_{\pm} = \{\Sigma \in {\cal A},\, det(C)
+ a b = \pm 1\}$. Then ${\cal A}$ is the disjoint union, ${\cal A}
= {\cal A}_{+} \,\dot\cup\, {\cal A}_-$. We shall show, ${\cal
A_+} = {\cal X}$, ${\cal A_-} = i {\cal X}$. We begin with the
observation that ${\cal X} \subset {\cal A_+}$. Indeed, $SU(4)$ is
connected, so ${\cal X}$ is connected. But, $\Sigma_c \in {\cal
X}$, and $det(C) + a b = 1$ for $\Sigma_c$. Therefore, $det(C) + a
b = 1$ for any $\Sigma \in {\cal X}$, so ${\cal X} \subset {\cal
A_+}$.

Now, take $\Sigma \in {\cal A_+}$. The condition $\Sigma
\Sigma^{\dagger} = 1$ implies, \beqn \label{unit1}C C^{\dagger} +
|a|^2 &=& 1\\\label{unit2}C^{\dagger} C + |b|^2 &=&
1\\\label{unit3} a^{*}\, \sigma_2 C^T \sigma_2 &=& b\,
C^{\dagger}\eeqn The remaining condition for $\Sigma \in {\cal
A_+}$ is, \beq \label{cond4} det(C) + a b = 1\eeq Equation
(\ref{unit1}) implies $|a| \leq 1$, and $C$ can be written
(non-uniquely) in the form, \beq \label{C} C = \sqrt{1-|a|^2} e^{i
\phi_c} S, \quad S \in SU(2)\eeq Substituting (\ref{C}) into
(\ref{unit2}), produces, $|a| = |b|$. We write, $a = |a| e^{i
\phi_a}$, $b = |b| e^{i \phi_b}$. If $a = 0$, the condition
(\ref{unit3}) is satisfied automatically, and (\ref{cond4})
implies, $e^{2 i \phi_c} = 1$. If $|a| = 1$, the condition
(\ref{unit3}) is again satisfied, and (\ref{cond4}) gives, $e^{i
(\phi_a + \phi_b)} = 1$. Finally, if $0 < a < 1$, (\ref{unit3})
becomes, \beq e^{2 i \phi_c} \sigma_2 S^{T} \sigma_2  = e^{i
(\phi_a + \phi_b)} S^{\dagger}\eeq But $SU(2)$ is pseudo-real:
$\sigma_2 S^{T} \sigma_2 = S^{\dagger}$, so $e^{2 i \phi_c} =
e^{i(\phi_a + \phi_b)}$. Substitution into (\ref{cond4}) gives
\beq e^{2 i \phi_c} = e^{i(\phi_a + \phi_b)} = 1\eeq We note that
if $e^{i \phi_c} = -1$, we can always reabsorb the negative sign
into the definition of $S \in SU(2)$. Thus, $\Sigma \in {\cal
A_+}$ if and only if it may be written as, \beq
\label{param}\Sigma = \left(\begin{array}{cc} |a| e^{i \phi_a}
\sigma_2 & \sqrt{1-|a|^2} S\\-\sqrt{1-|a|^2} S^{T} & |a| e^{-i
\phi_a} \sigma_2\end{array}\right),\quad |a| < 1, \quad S \in
SU(2)\eeq

We now show that any $\Sigma$ of form (\ref{param}) is in ${\cal
X}$. We let, $e^{i \xi} = \sqrt{1-|a|^2} + i |a|$. Define
matrices, $U_1, \,U_2, \,U_3, U \in SU(4)$ as, \beqn U_1 &=&
\exp\left(i \frac{\xi}{2}\left(\begin{array}{cc} 0 & - i
\sigma_2\\i \sigma_2 & 0\end{array}\right)\right)\\
U_2 &=& \left(\begin{array}{cc} -S & 0\\0 & 1\end{array}\right)\\
U_3 &=& \exp(i \phi_a B/2)\\\\U &=& U_3 U_2 U_1\eeqn One can now
check that $U \Sigma_c U^{T} = \Sigma$. This concludes the proof
of the fact that ${\cal X} = {\cal A}_+$. It is now trivial to
show that ${\cal A}_- = i {\cal A}_+ = i {\cal X}$.

The most practically useful result of the above discussion is the
explicit parametrization (\ref{param}) of the vacuum manifold
${\cal X}$. For instance, this parametrization allows one to prove
that the local minimum of the effective Lagrangian (\ref{Lch}) at
finite $\mu_B$ and $\mu_I$, originally constructed in
\cite{Splittorff:2000mm}, is, actually, global. Moreover, we can
now use the form (\ref{param}) to study the topology of the vacuum
manifold ${\cal X}$. Indeed, the matrix $S \in SU(2)$ can be
uniquely written as $S = X_0 + i X_i \sigma_i$, $X_0 X_0 + X_i X_i
= 1$. Also, $e^{i \phi_a} = X_4 + i X_5$, $X_4^2 + X_5^2 = 1$.
Clearly, ${\cal X}$ is in one to one correspondence with the set
of vectors in ${\mathbb R}^6$, $\sqrt{1-|a|^2}(X_0, X_1, X_2, X_3,
0, 0) + |a|(0,0,0,0,X_4, X_5)$, $0 \leq |a| \leq 1$. But this is
just a parametrization of $S^5$. Hence, ${\cal X} = SU(4)/Sp(4)
\cong S^5$.

Finally, let us discuss the $SU(4)$ transformation (\ref{rot}).
Since, $U_0 B U_0^{\dagger} = B$, $U_0 I U_0^{\dagger} = I$, the
kinetic part of the effective Lagrangian (\ref{Lch}) remains
invariant. Therefore, we have to discuss only the mass term: \beq
{\cal L}_m = -g Re \Tr({\cal M} \Sigma) = -g Re \Tr({\cal M} U_0
\tilde{\Sigma} U_0^T)\eeq where $\tilde{\Sigma}\in{\cal X}$ and,
therefore, can be written in the form (\ref{param}), as, \beq
\tilde{\Sigma} = \left(\begin{array}{cc} a \,\sigma_2 & C
\\-C^T & a^* \,\sigma_2\end{array}\right),\quad C = \sqrt{1-|a|^2}\, S, \;\;S\in SU(2)\eeq
Expanding, \beq {\cal L}_m = \frac{1}{2}g \left(e^{-i \theta/2}
\Tr(e^{i \alpha \sigma_3} \{M, C\}) + e^{i \theta/2} \Tr(e^{-i
\alpha \sigma_3} \{M, C^*\})\right)\eeq At this point, we again
use the pseudo-reality of $SU(2)$, $C^* = \sigma_2 C \sigma_2$,
obtaining after some algebra, \beq {\cal L}_m = 2 g m(\theta)
\Tr(C) = - g m(\theta) \Tr({\cal M}_0 \tilde{\Sigma})\eeq Thus, to
first order in $M$, all $\theta$ and $m_u, m_d$ dependence can be
incorporated into the quark mass $m(\theta)$ (\ref{m}).

\end{document}